%% file: main.tex
\documentclass[fleqn,10pt]{SelfArx}

\definecolor{color1}{RGB}{0,0,90}
\definecolor{color2}{RGB}{0,20,20}

\input{preamble_art}

\input{metadata_art}

\begin{document}

\flushbottom
\maketitle

\input{paper_body_core}

\bibliography{thesis}

\appendix
\input{appendix_a}

\end{document}

%% file: preamble_art.tex
\input{preamble_science}
\journalbuildtrue
\input{float_layout}

\usepackage{chngcntr}
\counterwithin{theorem}{section}

\usepackage{xurl}
\usepackage[hidelinks]{hyperref}
\input{preamble_cleveref}
\usepackage{adjustbox}

\providecommand{\keywordname}{Keywords}

%% file: preamble_science.tex
\newif\ifjournalbuild

\usepackage{tabularx}
\usepackage{array}
\usepackage{booktabs}

\usepackage{silence}
\usepackage[labelformat=simple]{subcaption}

\usepackage{textcomp}
\usepackage{gensymb}
\usepackage{mathtools}
\usepackage{amssymb}
\usepackage[multiple]{footmisc}

\usepackage[amsmath, thmmarks]{ntheorem}

\theoremstyle{nonumberplain}
\theoremheaderfont{\itshape}
\theorembodyfont{\normalfont}
\theoremseparator{.}
\theoremsymbol{\ensuremath{\square}}

\theoremstyle{plain}
\theoremheaderfont{\normalfont\bfseries}
\theorembodyfont{\itshape}
\theoremseparator{}
\theoremsymbol{}

\newsavebox{\codebox}
\usepackage{csquotes}

\usepackage[numbers,sort,compress,square]{natbib}
\usepackage[nottoc]{tocbibind}

\makeatletter
\def\NAT@spacechar{~}
\makeatother

\usepackage{tikz}
\usetikzlibrary{matrix,arrows.meta,positioning,fit,backgrounds,calc}

\usepackage{float}
\usepackage{graphicx}
\graphicspath{{./}}

%% file: float_layout.tex
\ifjournalbuild
  \usepackage{dblfloatfix}
  \usepackage{float}
\fi
\usepackage{adjustbox}
\usepackage{placeins}
\usepackage{longtable}
\usepackage{seqsplit}

\newcommand{\OceanTableSetup}{
  \footnotesize
  \setlength{\tabcolsep}{3pt}
  \renewcommand{\arraystretch}{1.05}
}

\newcommand{\OceanFigBegin}{
  \ifjournalbuild
    \begin{figure*}[!htbp]
    \centering
  \else
    \begin{figure}[htbp]
    \centering
  \fi
}

\newcommand{\OceanFigEnd}{
  \ifjournalbuild
    \end{figure*}
  \else
    \end{figure}
  \fi
}

\newcommand{\OceanTabBegin}{
  \ifjournalbuild
    \begin{table*}[!t]
    \centering
    \OceanTableSetup
  \else
    \begin{table}[htbp]
    \centering
    \OceanTableSetup
  \fi
}

\newcommand{\OceanTabEnd}{
  \ifjournalbuild
    \end{table*}
  \else
    \end{table}
  \fi
}

\newcommand{\OceanTabBeginCol}{
  \ifjournalbuild
    \begin{table}[htbp]
    \centering
    \OceanTableSetup
  \else
    \begin{table}[!htbp]
    \centering
    \OceanTableSetup
  \fi
}

\newcommand{\OceanTabEndCol}{
  \ifjournalbuild
    \end{table}
  \else
    \end{table}
  \fi
}

\newcommand{\OceanTabCaption}[1]{
  \caption{#1}
}

\newcommand{\OceanFigBeginCol}{
  \ifjournalbuild
    \begin{figure}[htbp]
    \centering
  \else
    \begin{figure}[!htbp]
    \centering
  \fi
}

\newcommand{\OceanFigEndCol}{
  \ifjournalbuild
    \end{figure}
  \else
    \end{figure}
  \fi
}

\newcommand{\OceanFigCaption}[1]{
  \caption{#1}
}

\newcommand{\OceanResizeTable}[1]{
  \begin{adjustbox}{max width=\linewidth}
    #1
  \end{adjustbox}
}

%% file: preamble_cleveref.tex
\usepackage[capitalise]{cleveref}
\makeatletter
\DeclareRobustCommand{\@@number}[1]{#1}
\makeatother

\Crefname{observation}{Observation}{Observations}
\Crefname{assumption}{Assumption}{Assumptions}

%% file: metadata_art.tex
\JournalInfo{Preprint}
\Archive{0.1}

\PaperTitle{Stylometric Defenses Against Author Impersonation in Software Repositories}

\Authors{Ravich Leonid\textsuperscript{1} and Fire Michael\textsuperscript{1}}
\affiliation{\textsuperscript{1}\textit{Ben-Gurion University of the Negev, Computer Science Department}}

\Keywords{code authorship verification, software supply chain, stylometry, contrastive learning, anomaly detection}

\input{abstract_art}

%% file: abstract_art.tex
\Abstract{\input{abstract_body}}

%% file: abstract_body.tex
Software supply-chain attacks increasingly exploit an identity gap where compromised maintainer accounts authorize 
malicious changes. This work evaluates \emph{patch-level} authorship verification as a behavioral defense layer, 
showing that stylometric analysis can operate not only on full source files but also on patch-level commits. We fine-tune a 
cross-modal transformer on more than 20 years of Linux kernel commit history to embed code diffs and commit messages 
into a unified stylometric space, achieving ROC AUC of 0.93 for open-world authorship verification. 
We then use these representations in a streaming anomaly detector suited to continuous integration and deployment (CI/CD) 
settings. We validate the pipeline on two retrospective supply-chain incidents involving different patch characteristics: 
the 2021 PHP backdoor and the 2026 ForceMemo/GlassWorm campaign. Without retraining, the proposed detector 
surfaces both PHP forged commits within approximately 1\% of the maintainer audit queue and ranks the 28 scoreable 
ForceMemo spoofs with a median per-repository review burden of 0.8\%. 
These results indicate that cross-modal patch-level embeddings can support behavioral triage against 
author impersonation in real-world repositories.

%% file: paper_body_core.tex
\section{Introduction}
\input{introduction}

\section{Related Work}
\label{sec:related}
\input{related_work}

\section{Methodology}
\label{sec:methods}
\input{methods}

\subsection{Experimental Setup}
\label{sec:experiments}
\input{experiments}

\section{Results}
\label{sec:results}
\input{results}

\section{Discussion}
\label{sec:discussion}
\input{discussion}

\section{Conclusions}
\label{sec:conclusions}
\input{conclusions}

\section*{Acknowledgments}
While drafting this article, we used Cursor and Claude for editing and for software development supporting the experiments.

%% file: introduction.tex
    The global software supply chain relies on a fragile assumption of identity. While Open Source Software (OSS) forms the foundation of modern digital infrastructure, 
    its decentralized nature introduces systemic vulnerabilities. Existing defenses focus on artifact integrity and access controls, 
    such as Software Bills of Materials (SBOM) and multi-factor authentication (MFA)~\cite{souppaya_secure_2022}. 
    However, these measures leave a critical internal identity gap. Cryptographic signatures and access controls attest to who authorized a change, not who actually 
    wrote the code. As shown by~\citet{Siadati2024}, maintainer accounts can be compromised via social engineering or stolen credentials.
    Once that occurs, attackers can push malicious patches under the victim's identity and thereby bypass traditional security gates~\cite{Williams2025}. 
    Recent high-profile exploits demonstrate how attackers exploit this gap to propagate malware downstream. The 2021 PHP source-code backdoor~\cite{Popov2021PHPIncident,PHPWatch2021}, the 2024 XZ Utils compromise~\cite{CISA2024XZ}, and the 2026 ForceMemo/GlassWorm campaign~\cite{ForceMemoGlassWorm2026} show the devastating impact of injected commits under trusted identities. 
    These ecosystem-level campaigns, particularly the ForceMemo/GlassWorm activity examined in our incident validation, further illustrate that this risk extends beyond single 
    repositories into coordinated abuse of developer personas. To counter this, security models need a behavioral layer that verifies identity based on the fingerprint 
    of the work itself.
    Code stylometry, the quantitative analysis of individual writing styles in source code, offers a content-centric complement to traditional provenance checks. 
    Yet a significant gap exists between theoretical stylometry and practical application. As demonstrated by~\citet{Caliskan2015}, models can achieve high accuracy in controlled environments like Google Code Jam. However, \citet{Dauber2019GitBlameWho} showed that this performance degrades significantly in incremental, real-world repositories, a finding further supported under realistic Git evaluation settings by~\citet{bogomolov2021authorship}. 
    Furthermore, existing evaluations typically operate at the file or function level, whereas supply chain impersonation often involves small, 
    atomic patch-level edits.

    This motivates two empirical research questions, evaluated in Section~\ref{sec:discussion}.
    \textbf{RQ1 (patch-level authorship signal):} Do real-world version-control patches contain enough behavioral signal to support 
    authorship verification for developers unseen during training (open-world)?
    \textbf{RQ2 (operational misuse detection):} Can this signal be used in an anomaly detector to surface commits that may 
    reflect author impersonation or account compromise?

    To address these questions, we treat the patch and the commit message as a behavioral artifact. 
    We develop a cross-modal supervised contrastive learning framework that fine-tunes UniXcoder~\citep{guo2022unixcoder} to jointly encode commit messages and unified code diffs into a normalized stylometric embedding space. 

    We train and evaluate this model on a massive twenty-one-year slice of the Linux kernel Git repository. Building upon these learned embeddings, 
    we deploy a sliding-window anomaly detector to monitor stylistic trajectories and flag deviations. 
    
    Our evaluation shows that patch-level behavioral signals can support both open-world verification and 
    ranked anomaly triage. 
    In the Linux kernel benchmark, the early-fusion UniXcoder model achieves 0.9322 ROC AUC for patch-level authorship verification,
    and the corresponding detector reaches 0.939 AUPRC under synthetic author swaps.
    In external incident checks without retraining, our proposed pipeline ranked both 2021 PHP forged commits within the top 1.03\%
    of a retrospective audit queue, and ranked the typical ForceMemo/GlassWorm spoof after a median of one benign commit (0.8\% of the per-repository scored queue across 28 scoreable spoofs). 
    Under maintainer-scoped swap, the anomaly detector's late-fusion rules typically surface injected commits within roughly the top 0.1\% of the review queue
    (median 50 to 65 false positives before the injected commit among approximately 60{,}000 commits).
    These results position stylometric scoring as a behavioral triage layer for supply-chain review, complementing signatures, access controls, and human code review. 
    The overall pipeline is illustrated in Figures~\ref{fig:training_pipeline} and~\ref{fig:anomaly_detection}.

    \OceanFigBegin
      \includegraphics[width=\textwidth]{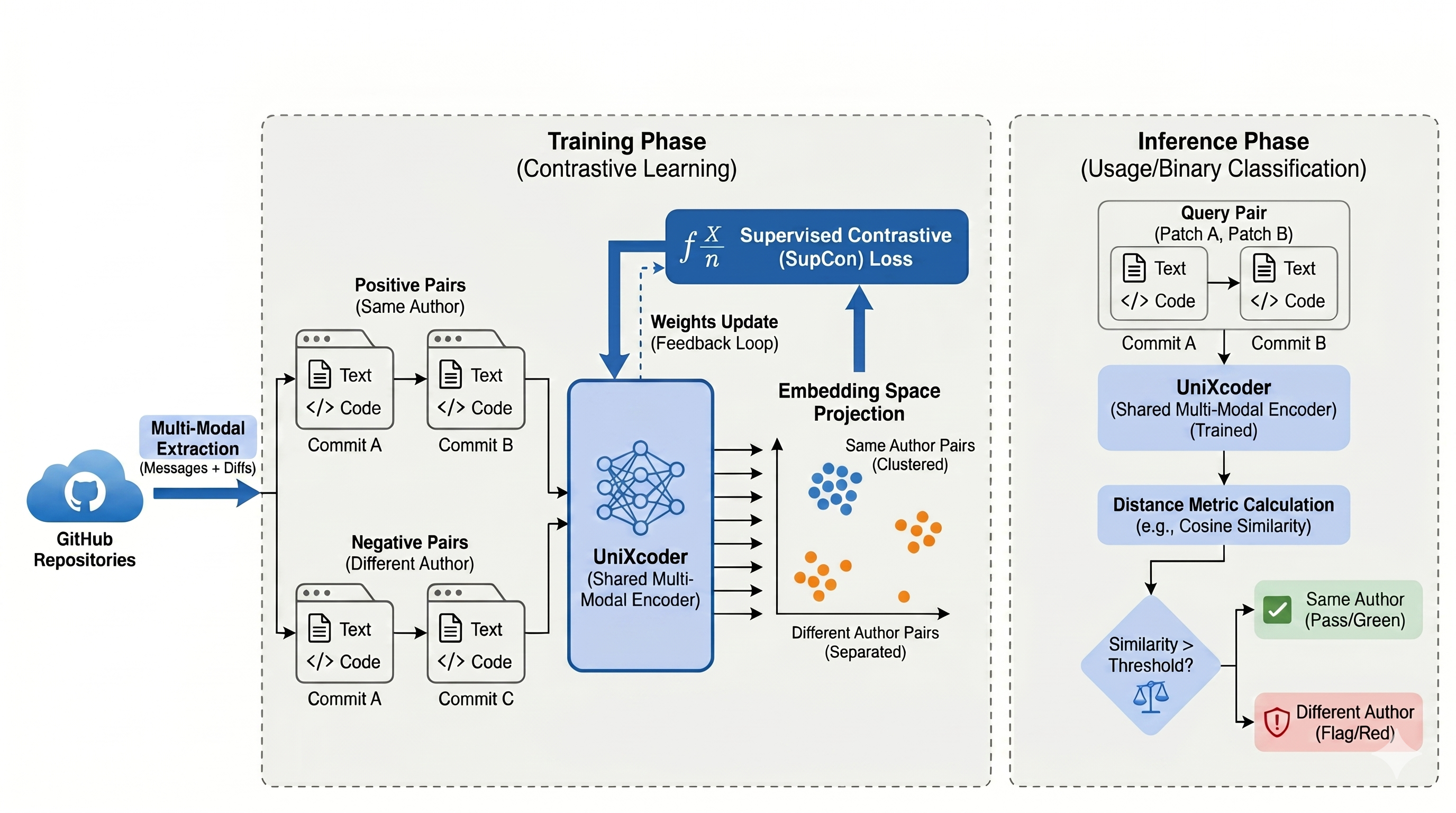}
      \caption{Authorship verification training pipeline.}
      \label{fig:training_pipeline}
    \OceanFigEnd

    \OceanFigBegin
      \includegraphics[width=\textwidth]{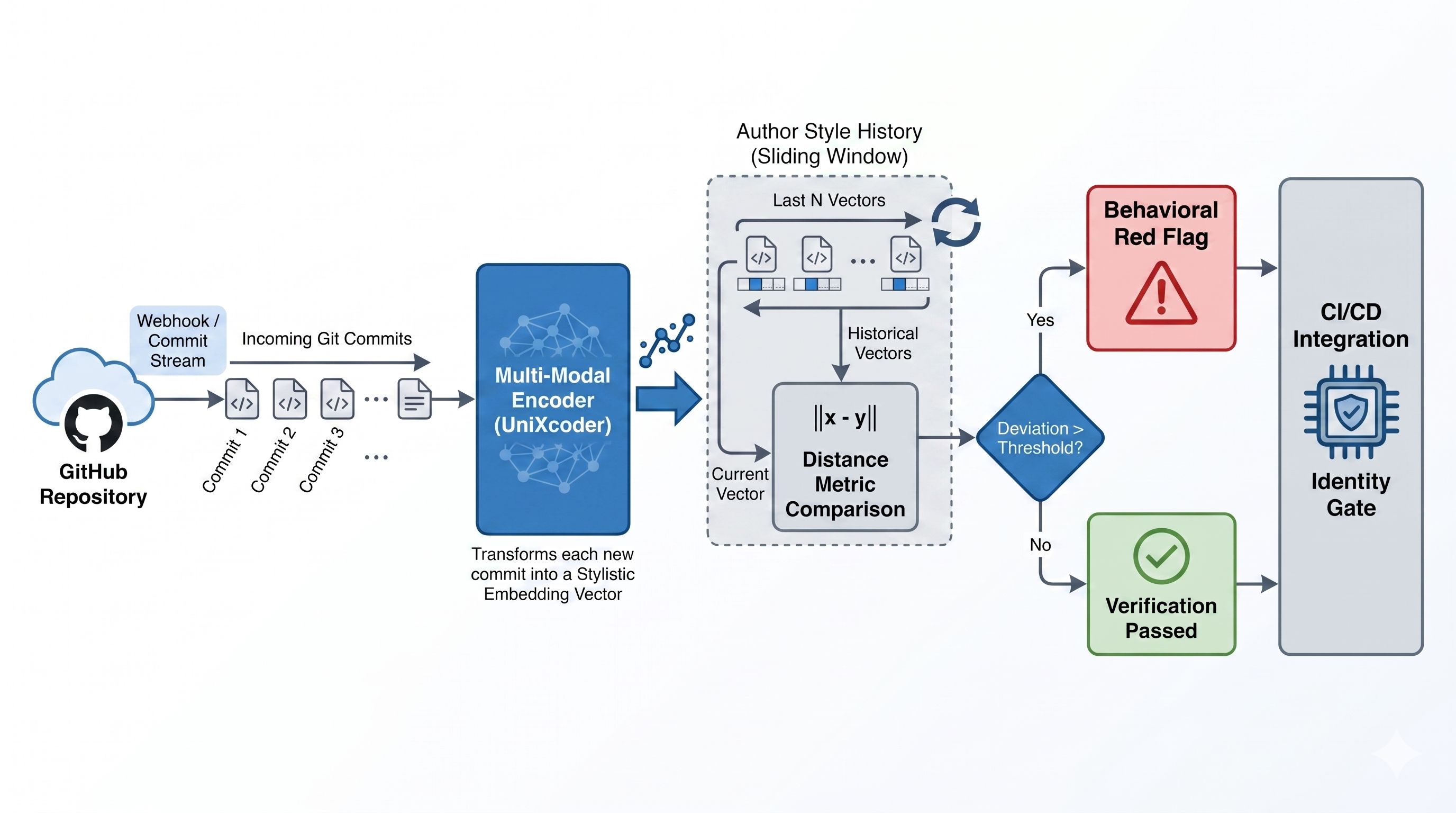}
      \caption{Anomaly detection pipeline.}
      \label{fig:anomaly_detection}
    \OceanFigEnd

    This study's key contributions are threefold:
    \begin{enumerate}
        \item \textbf{Large-scale open-world patch-level benchmark:} We construct a reproducible binary authorship-verification dataset from over two decades of Linux kernel history (398{,}505 commits, 4{,}474 developers). Using strict author-disjoint splits, this benchmark bridges the gap between controlled stylometry and noisy, incremental OSS development.
        \item \textbf{Cross-modal contrastive evaluation and modality analysis:} We introduce a supervised contrastive learning framework that jointly encodes commit messages and unified diffs. By evaluating both early-fusion and independent modalities, we characterize the modality hierarchy at the patch scale, demonstrating that natural-language messages drive average verification performance. At the same time, code acts as a necessary anchor for detection.
        \item \textbf{Anomaly detection for code repositories:} We build a sliding-window detector with patch-size-gated cross-modal fusion and validate it on synthetic kernel benchmarks and real supply-chain incidents (PHP and ForceMemo/GlassWorm), providing a continuous behavioral triage layer for CI/CD environments.
    \end{enumerate}
    
    \ifjournalbuild
    The remainder of this article is organized as follows: \Cref{sec:related} reviews the literature on code authorship and supply chain security. \Cref{sec:methods} presents our methodology, with \Cref{sec:experiments} detailing the experimental setup. \Cref{sec:results} reports the empirical findings, \Cref{sec:discussion} interprets them, and \Cref{sec:conclusions} summarizes implications and future work.
    \else
    The remainder of this thesis is organized as follows: \Cref{sec:related} reviews the literature on code authorship and supply chain security. \Cref{sec:methods} presents our methodology, with \Cref{sec:experiments} detailing the experimental setup. \Cref{sec:results} reports the empirical findings, \Cref{sec:discussion} interprets them, and \Cref{sec:conclusions} summarizes implications and future work. An appendix provides supporting implementation details.
    \fi

%% file: related_work.tex
The increasing frequency of software supply-chain compromises \citep{Ladisa2023,ohm2020backstabber} 
has motivated research on attacks and mitigations in open-source ecosystems. 
We review supply-chain defenses, classical code authorship verification, 
and recent contrastive approaches to 
position patch-level behavioral verification as a complementary defense layer.

\subsection{Supply-Chain Attacks and the Identity Gap} \label{sec:supply-chain-identity-gap}
Static code analysis detects malicious logic without executing the code. \citet{ragkhitwetsagul2018comparison} compared code similarity analyzers on copy-and-modify scenarios, 
including obfuscated clone pairs, and reported which token, text, tree, and graph-based measures still recover modified matches. 
Their findings indicate that static similarity detection benefits from structural and feature-rich comparators rather than isolated 
token overlap. \citet{Garrett2019Suspicious} addressed supply-chain risks by proposing methods to detect suspicious package 
updates in software ecosystems. 
By analyzing metadata and static features of new releases, they identified potentially malicious updates before they were widely adopted. 
To improve scalability, \citet{Alomari2019Scalable} introduced a control-flow-graph approach for detecting source code similarity 
across large repositories, enabling efficient identification of cloned or near-duplicate segments in extensive codebases.

Complementary signature-based approaches refine malware-detection rules without executing target code. \citet{naik2020evaluating} evaluated automatically generated YARA rules and proposed fuzzy-hash refinements that improve rule effectiveness.
Recent static analysis research emphasizes deep-learning models trained on large code corpora. For instance, \citet{Tsfaty2023MSDT} introduced MSDT, a transformer-based method that identifies anomalous functions with precision@$k$ up to 0.91 on some function types. Despite these advances, static techniques are limited by high false-positive rates and a lack of resistance to obfuscation. Consequently, recent studies have adopted broader, defense-oriented frameworks. \citet{AbuIshgair2024AStRA} developed the AStRA model. This graph-based framework maps security objectives to defense techniques by capturing causal relationships among artifacts, steps, resources, and principals in the supply chain. These systematizations underscore the necessity for holistic approaches rather than isolated attack taxonomies.
 
Dynamic analysis addresses the limitations of static methods by executing code in sandboxed environments to monitor system calls, dependency usage, and runtime behavior. Early approaches, such as those reviewed by \citet{Idika2007Survey}, observed system calls, variable values, and input/output during execution to detect malicious behavior. In addition to academic prototypes, practical pipelines have been developed. For example, \citet{Zheng2024OSCAR} fully executed NPM and PyPI packages in a sandbox, fuzzed exported functions, and monitored API-level behaviors, achieving high F1 scores and significantly reducing false-positive rates compared to earlier tools.
 
Additional approaches model sequences of malicious behaviors. For example, \citet{Zhang2025Cerebro} encode high-level behavior patterns into sequences and fine-tune a pre-trained language model, enabling the detection of malicious packages across different ecosystems. In real-world deployments, they  identified 683 malicious PyPI packages and 799 malicious NPM packages.
 
\citet{fleshman2019evading} demonstrated that attackers can evade machine-learning malware classifiers, highlighting the need for continuously evolving defensive techniques. Large-scale analyses of code contribution practices reveal systemic vulnerabilities. For example, \citet{Holtgrave2025} studied 50,328 critical open-source projects. They found that contribution workflows are exploitable in 85.9 percent of projects, enabling attackers to hijack over 573,000 email addresses and forge historical contributions. Notably, cryptographic commit signing remains rare. Only 2 percent of users sign all their commits~\citep{Holtgrave2025}. Even when signatures are present, compromised keys or insider misuse mean that signatures guarantee provenance but not authorship. These findings emphasize the necessity of identity-aware defenses to complement traditional access controls. Although current methods show that malicious commit detection is possible, they also reveal an arms-race dynamic: as detection methods advance, attackers adapt by modifying obfuscation strategies or imitating benign functionality. Both static and dynamic analyses focus on the nature of code alterations but do not address the identity of contributors. This identity gap necessitates behavioral defensive measures, which are discussed in the following subsection.

\subsection{Classical Code Authorship Verification} \label{sec:classical-authorship-verification}
Early code stylometry methods used handcrafted features such as token n-grams, keyword frequencies, indentation patterns, and abstract syntax tree traversals, combined with classifiers like support vector machines or logistic regression. \citet{Caliskan2015} reported high attribution accuracy on Google Code Jam submissions, while \citet{wang2018integration} introduced feature variations and hybrid classifiers that also reported strong performance on Python tasks. A related but distinct thread of natural-language authorship verification, often evaluated in shared tasks like PAN~\citep{stamatatos2023pan}, has shown that textual habits carry strong identity signals, though such signals can be confounded by topic variations~\citep{seroussi2014topicmodels}. Beyond file-level stylometry, later work investigated authorship attribution on smaller and noisier code fragments.

\citet{Dauber2019GitBlameWho} presented one of the first studies to examine authorship attribution for short, incomplete source-code snippets extracted directly from version-control histories. 
The authors constructed a dataset of partial and uncompileable fragments and evaluated an ensemble of classical classifiers trained on lexical, syntactic, and structural metrics, including token $n$-grams, indentation statistics, and AST node distributions. 
Their ensemble approach achieved competitive accuracy on fragment-level attribution by averaging predictions across multiple small samples, but also revealed that accuracy drops sharply in real-world repositories or when the stylistic signal is sparse. 
 
However, classical approaches generally assume closed-set identification and 
are evaluated at the file level. 
Their reliance on handcrafted features limits scalability, and accuracy often degrades considerably under open-world scenarios with previously unseen authors. 
For example, small numbers of files per author can significantly reduce attribution performance \citep{abazari2023dataset}, and adversarial or semantics-preserving edits can mislead classifiers and collapse attribution accuracy \citep{Quiring2019Misleading, Abuhamad2025SHIELD}.

\subsection{Pre-Trained Models and Contrastive Learning} \label{sec:pretrained-contrastive}
Previous neural authorship systems abandoned handcrafted metrics in favor of learned representations.  \citet{yang2017authorship} proposed one of the first neural approaches by computing 19 lexical, layout, structural, and syntactic metrics and feeding them into a back-propagation neural network tuned via particle swarm optimization; on 3,022 Java files from 40 authors, their model reported strong attribution accuracy and outperformed earlier heuristic methods.  Building on this idea, \citet{abuhamad2019cnn} moved from manually engineered metrics to learned features by combining TF--IDF representations, word embeddings, and convolutional networks.  Their CNN-based system achieved very high attribution accuracy on Google Code Jam cohorts and strong generalization across GitHub repositories in C and C++ code.  Such convolutional architectures learn discriminative code embeddings across languages but still treat authorship as a closed-set classification problem.

To address open-set verification, \citet{white2021deep} formulated authorship as a metric-learning task.  Their model utilizes a bidirectional LSTM and a contrastive loss, ensuring that code fragments written by the same author map to nearby points in an embedding space. In contrast, fragments from different authors are positioned farther apart.  Classification of these embeddings with both support-vector machine and k-nearest neighbors classifiers achieved strong accuracy, demonstrating that metric learning can support both attribution and verification.  However, the approach still requires substantial training data and shows limited generalization to real-world repositories.

Recent work explores adversarial and contrastive frameworks that explicitly disentangle stylistic signals from functionality.  \citet{ou2023scsgan} proposed \emph{SCS-GAN}, 
an adversarial model that uses multi-head attention to isolate stylistically informative tokens and trains a generator–discriminator pair to produce functionality-agnostic stylometric representations.  
Evaluated on four out-of-sample datasets from a real-world programming competition, SCS-GAN consistently outperformed previous representation models, demonstrating the promise of adversarial/contrastive techniques for robust authorship verification.

Despite many advances, most neural authorship models are trained on contest-style datasets such as Google Code Jam and are evaluated in closed, monolithic settings. When applied to small fragments or real-world open-source repositories, their performance often degrades significantly. \citet{Dauber2019GitBlameWho} showed that extending contest-based attribution techniques to incomplete source code fragments in collaborative repositories is substantially more difficult than file-level attribution. Similarly, \citet{bogomolov2021authorship} demonstrated that models achieving high accuracy on benchmark datasets experience severe drops when evaluated on more realistic Git histories. \citet{Abuhamad2021LargeScale} advanced large-scale code authorship identification with deep feature learning, yet adversarial studies such as \citet{li2022ropgen} and \citet{Quiring2019Misleading} revealed that minor semantic-preserving edits can mislead stylometric classifiers. These findings suggest that contest-trained neural models struggle to capture subtle stylistic cues in small edits or diverse collaborative environments, motivating current work on pre-trained code models and contrastive learning strategies that can learn reusable stylometric embeddings across languages and tasks.

Recent advances in pre-trained code representation models have substantially improved downstream performance across tasks such as code summarization, clone detection, and defect prediction.  
Among the most influential architectures are \emph{CodeBERT}~\citep{feng2020codebert}, \emph{GraphCodeBERT}~\citep{guo2021graphcodebert}, \emph{CodeT5}~\citep{wang2021codet5}, and \emph{UniXcoder}~\citep{guo2022unixcoder}.
Commit-specific extensions of this paradigm include \emph{CommitBERT}~\citep{jung-2021-commitbert}, which fine-tunes CodeBERT on commit message--code diff pairs to improve commit message generation, and \emph{CommitBART}~\citep{Liu2022CommitBART}. This encoder-decoder model further extends this by conditioning generation on structured patch representations.
While CommitBERT and CommitBART target commit-message generation rather than authorship, their joint modeling of commit messages and code diffs is closely related to the cross-modal input representation used in this work.
CodeSage further scales this family of code representation models by training on large code corpora with architectures designed for broad code-understanding tasks~\citep{CodeSage2024}.

UniXcoder, in particular, integrates both unimodal and cross-modal learning objectives to jointly encode code, natural language, and their interrelations.  
It employs a unified transformer architecture pre-trained on large-scale parallel corpora combining source code and natural language documentation across multiple programming languages.  
By introducing task-specific prefixes and modality embeddings, UniXcoder supports diverse generation and understanding tasks, including code completion, summarization, and retrieval, within a single model.  
Its bidirectional encoder and causal decoder training objectives allow it to capture both local syntactic cues and long-range semantic dependencies.  
These properties make UniXcoder a plausible encoder for stylometric analysis of commits, where both natural-language intent and code-change structure may carry author-specific signal.

Building on the success of such learned representations, recent systems have applied contrastive learning directly to authorship 
verification tasks. OCEAN~\citep{machtle2024ocean} applied supervised contrastive learning to binaries and reported strong open-world 
verification performance, while CLAVE~\citep{alvarez2025clave} reported very high AUROC on Python benchmarks. 
These results highlight the power of embedding-based verification but also reveal key gaps. 
First, most validations depend on GCJ, whose contest submissions lack the collaborative realism of version-control repositories~\citep{abazari2023dataset}. Second, most experiments operate at the file granularity \citep{ou2023scsgan,alvarez2025clave,abuhamad2019cnn,white2021deep,yang2017authorship,hozhabrierdi2020zeroshot,wang2018integration,Caliskan2015} or at the function granularity \citep{machtle2024ocean,bogomolov2021authorship}. \citet{Dauber2019GitBlameWho} studied short, incomplete code snippets extracted from version-control histories, and showed that attribution for such fragments is substantially harder than file-level attribution. However, that work targets the closed-set attribution task on small corpora rather than open-world binary verification. To our knowledge, no prior study has performed author-disjoint open-world binary verification on unified Git patches at Linux-kernel scale, combining joint commit-message and diff encoding with maintainer-stream anomaly scoring.

This study extends previous research by applying patch-level authorship verification to real open-source software projects. The evaluation determines whether contrastively fine-tuned transformer models can provide a useful stylometric signal for author identity verification under credential-theft and insider-misuse scenarios.

%% file: methods.tex
This section defines the methodology used for converting repository history into patch-level
stylometric signals. We first formalize binary authorship verification and streaming anomaly
detection, then describe corpus construction, sample preparation, representation learning,
modality-aware score fusion, and evaluation protocols.

\subsection{Task Formulation}\label{sec:problem_definition}

\textbf{Authorship verification.} As formalized by~\citet{Koppel2014} and operationalized in shared evaluations like PAN~\citep{stamatatos2023pan}, authorship verification is traditionally defined as a binary decision problem. That is, given a known document and a questioned document, the task is to determine whether the two samples were written by the same author. Following this standard formulation, 
we instantiate the task at the patch level: given two commit samples $c_i$ and $c_j$, 
decide whether they share the same author.
This differs from author attribution (closed-set), where each document is assigned to one 
of a fixed set of known authors~\citep{Koppel2014}. 
Author verification is open-world in the sense that test-time authors might not appear in training, 
which supports impersonation screening when the impostor is not enumerated at training time. 
We learn $\mathcal{F}\colon c \mapsto \mathcal{F}(c) \in \mathbb{R}^d$ and score pairs by cosine similarity,
\[
\mathrm{sim}(c_i,c_j)=\frac{\mathcal{F}(c_i)\cdot\mathcal{F}(c_j)}{\|\mathcal{F}(c_i)\|\,\|\mathcal{F}(c_j)\|}.
\]
A pair is predicted as \emph{same-author} if $\mathrm{sim}(c_i,c_j)>\tau$ for an operating threshold $\tau$. 
Several prior works perform open-world source code authorship verification by mapping code samples to embeddings and scoring pairs 
using cosine similarity (e.g., \citep{white2021deep,machtle2024ocean,alvarez2025clave}).
These approaches are discussed in Section~\ref{sec:pretrained-contrastive}.

\textbf{Anomaly detection.} We reuse $\mathcal{F}$ and express dissimilarity as cosine distance $d = 1 - \mathrm{sim}$.
When simulating the detector, commits from a given author identity are processed in chronological order. Each new commit is compared only with that specific identity's earlier commits.
The detector, therefore, asks whether a new patch is unusually distant from the identity's recent history under a thresholded anomaly score.

\subsection{Data Preparation}\label{sec:data_preparation}

Prior neural models for code authorship verification have often relied on contest-style monolithic submissions, such as Google Code Jam~\citep{abuhamad2019cnn,ou2023scsgan,shi2024transformer}.
Such settings capture limited incremental editing compared with long-running Git projects~\citep{abazari2023dataset,Dauber2019GitBlameWho,bogomolov2021authorship}. 
We therefore construct samples from active open source repositories. 
For each commit, we extract the author identity, timestamp, commit message, and unified code diff using PyDriller~\citep{Spadini2018PyDriller}, and store the extracted records in DuckDB~\citep{Raasveldt2019DuckDB}.
We retain only authors with enough commits for evaluation, then partition the data into training, validation, and test sets.

\textbf{Author identity.} Author labels are taken from version-control metadata.
Because a developer may commit under several name spellings and email addresses over time~\citep{Bird2006,Kouters2012},
we merge commit records that share a normalized email or a normalized display name into a single person-level identity.
The data splits are formed over these resolved identities so that no person appears in more than one split, ensuring an open-world evaluation on unseen authors.

\textbf{Sample definition.}\label{def:sample} A \emph{sample} is a single commit that modifies at
least one source file according to the specified language or path criteria, 
resulting in a unified diff. Each sample encapsulates a sanitized commit message alongside a 
normalized patch string. To preserve the commit as the atomic unit of analysis and to prevent the artificial splitting of 
correlated changes, multi-file modifications within a single commit are aggregated into a single sample.
These are concatenated using explicit structural markers, such as \texttt{<FILE>} and \texttt{<HUNK>}. 
This follows CC2Vec~\citep{CC2Vec2020}, which represents code changes via their hierarchical patch structure and shows that such representations are useful across software change tasks.

\textbf{Preprocessing.} Preprocessing deterministically maps each retained commit to a textual representation derived from commit metadata, 
the commit message ($t_{\mathrm{msg}}$), and the code diff ($t_{\mathrm{code}}$). Metadata such as the author's identity and commit timestamp are used only 
for supervision and chronological ordering. The encoder input is produced through message cleaning, structural patch linearization, 
content normalization, and final sequence assembly:

\begin{enumerate}[label=\textbf{Step \arabic*:}, leftmargin=*]
    \item \textbf{Message Cleaning.} Let $t_{\mathrm{msg}}$ be the raw commit message. We remove structured or potentially identifying artifacts that are not 
    part of the developer's free-form phrasing, including email addresses, common trailer fields (e.g., sign-off and review tags), 
    URLs, and issue references. The result is a cleaned message $t'_{\mathrm{msg}}$.

    \item \textbf{Structural Patch Linearization.} Let $t_{\mathrm{code}}$ be the raw code diff. We transform the diff into a single ordered stream $t'_{\mathrm{code}}$, where
    each line is annotated with its role in the patch, such as added (\texttt{<ADD>}), deleted (\texttt{<DEL>}), or context (\texttt{<CTX>}) content. 
    Additional markers such as \texttt{<FILE>} and \texttt{<HUNK>} preserve file- and hunk-level structure within a commit, following CC2Vec~\citep{CC2Vec2020} and PatchNet~\citep{Hoang2019PatchNet}, which model code changes as structured patch representations.

    \item \textbf{Content Normalization.} To reduce sensitivity to volatile surface forms while preserving structural information, 
    we normalize selected literals in the linearized patch $t'_{\mathrm{code}}$ to produce the final code representation $t''_{\mathrm{code}}$. In particular, hexadecimal constants, numeric literals, and web addresses are mapped to 
    stable placeholders such as \texttt{\_\_HEX\_\_}, \texttt{\_\_NUM\_\_}, and \texttt{\_\_WEB\_\_}. 
    This follows program-normalization work by Wang et al.~\citep{Wang2020ProgramNormalization}, which reduces superficial coding variation while preserving authorship-relevant structure.

    \item \textbf{Joint Sequence Assembly (Early Fusion).} For cross-modal models, the final input sequence $S$ is formed by concatenating the cleaned commit message and the normalized patch representation ($S = t'_{\mathrm{msg}} \parallel t''_{\mathrm{code}}$). This joint encoding allows the model to capture relationships between natural-language intent and code-change content,
    following commit-level models such as PatchNet~\citep{Hoang2019PatchNet}, CommitBART~\citep{Liu2022CommitBART}, and CommitBERT~\citep{jung-2021-commitbert}, which represent commit messages together with code changes. Single-modality experiments bypass this step and use either $t'_{\mathrm{msg}}$ or $t''_{\mathrm{code}}$ directly.

\end{enumerate}

Steps~1--3 define the modality-specific representations, while Step~4 defines the combined input for early fusion. Figure~\ref{fig:preprocessing_example} illustrates the transformation from raw commit fields to model input.

\OceanFigBegin
\includegraphics[width=\textwidth]{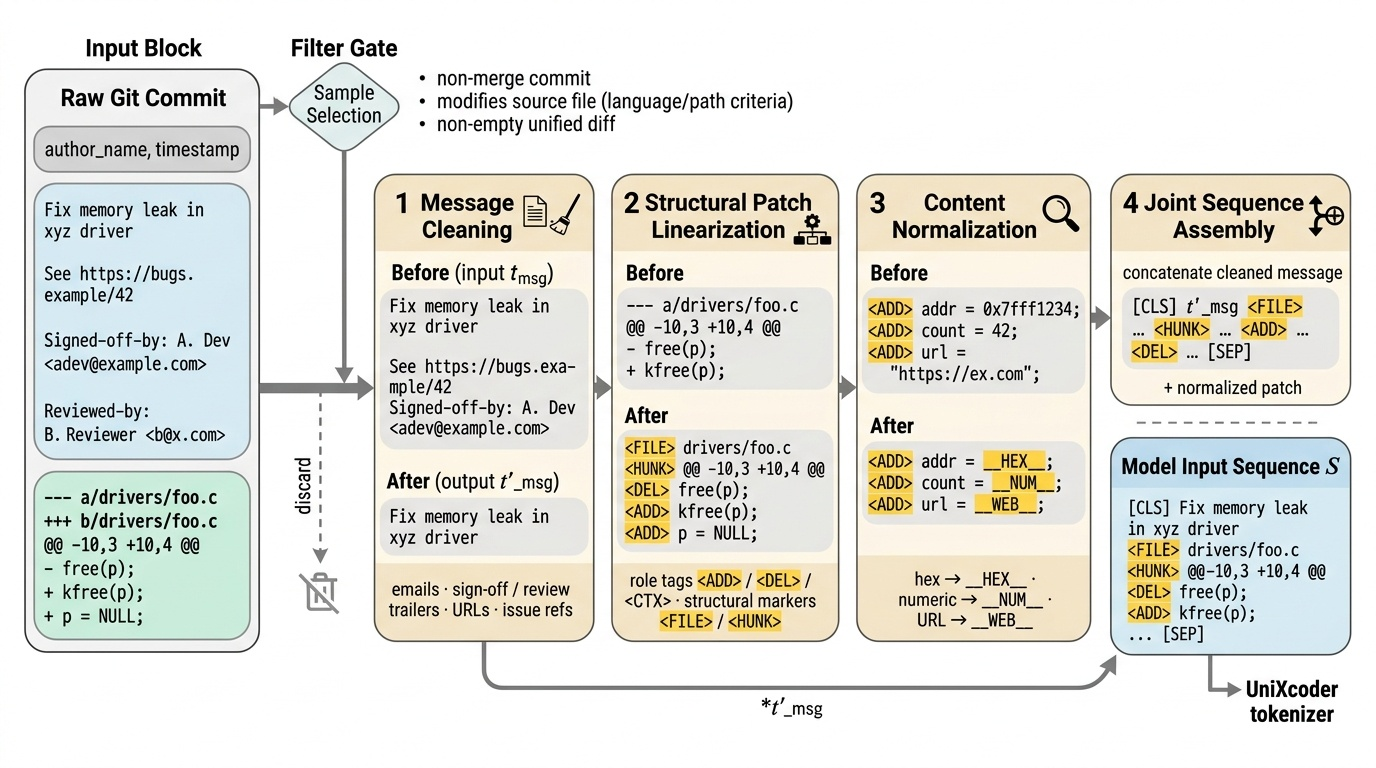}
\caption{Commit preprocessing example.}
\label{fig:preprocessing_example}
\OceanFigEnd

\subsection{Representation Learning}\label{sec:training}

To learn the embedding space $\mathcal{F}$, we fine-tune a pre-trained encoder backbone using the supervised contrastive (SupCon) loss~\citep{Khosla2020supcon}, as previously applied to multi-author writing style analysis~\citep{ye2023supcon}.
This encoder-plus-contrastive recipe also aligns with recent neural code stylometry systems such as OCEAN~\citep{machtle2024ocean} and CLAVE~\citep{alvarez2025clave}.
Figure~\ref{fig:representation_learning_pipeline} illustrates our representation learning pipeline.

\textbf{Sampling strategy.} Contributor activity in open-source projects follows a heavy-tailed distribution. 
Randomly sampling individual commits over-represents prolific authors and often yields mini-batches lacking the positive pairs required for contrastive learning. 
To address this, we employ \emph{unit-based sampling}. We group $u$ distinct commits from the same author into a single unit, constructing training batches by concatenating these shuffled units. 
By setting $u \ge 2$, we guarantee that every anchor commit has at least one positive match within the batch.
This construction satisfies the batch-positive requirements of the SupCon loss~\citep{Khosla2020supcon}.
Unit-based sampling also mitigates the effects of commit-frequency imbalance.

\textbf{In-domain batch construction.} As in natural-language authorship verification, shared topic vocabulary can confound stylistic signal~\citep{seroussi2014topicmodels,altakrori-etal-2021-topic-confusion}.
In collaborative repositories, developers often work repeatedly within specific subsystems, so diffs from different authors may share topical vocabulary and APIs~\citep{ou2023scsgan}. 
If negative examples are drawn from unrelated areas, the model might learn to distinguish domains rather than actual author styles. 
To prevent this shortcut, we optionally restrict each mini-batch to commits from a single coarse \emph{domain}, derived from path-to-subsystem mappings. 
Domain labels are used strictly for batching and are never provided as input to the encoder. 
This ensures that negative examples share topical overlap, forcing the model to learn true stylometric differences rather than domain boundaries.

\textbf{Objective function.} We optimize the supervised contrastive loss of Khosla et al.~\citep{Khosla2020supcon} using 
author identities as supervised class labels. During training, this objective forces embeddings of commits by the same author to cluster closely while pushing apart embeddings of commits from different authors within the batch.

\OceanFigBegin
\includegraphics[width=\textwidth]{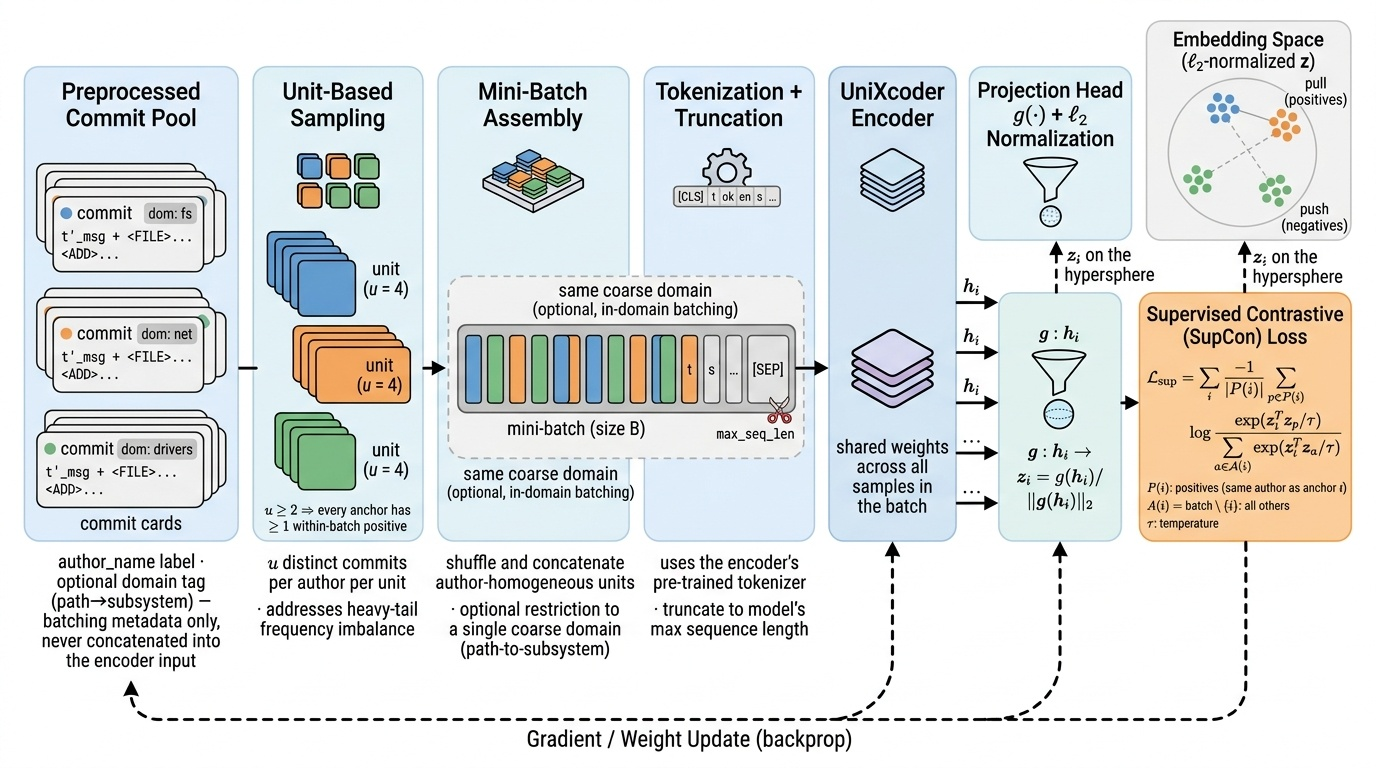}
\caption{Representation learning pipeline.}
\label{fig:representation_learning_pipeline}
\OceanFigEnd

\subsection{Late Fusion}\label{sec:score_fusion}

To capture stylometric signals from both commit messages and code changes, models typically employ early fusion, concatenating raw modalities into a single sequence before encoding~\citep{Hoang2019PatchNet,Liu2022CommitBART,jung-2021-commitbert}. 
While this allows the model to learn deep, fine-grained interactions between natural language and code, 
it forces the network to process both modalities simultaneously. In the context of supply-chain attacks, 
this coupling can be a disadvantage, as malicious commits are often highly asymmetric. 
A targeted attack might consist of a tiny payload that carries almost no stylometric signal in the code diff, while still containing a usable, deceptive commit message. 
To address this variable reliability, we evaluate a late fusion approach, which processes modalities through independent encoders and integrates information at the decision level~\citep{Atrey2010MultimodalFusion,Baltrusaitis2019MultimodalSurvey}. 
This architecture yields two parallel anomaly score streams, $s_t^{\mathrm{code}}$ and $s_t^{\mathrm{text}}$, decoupling the modalities to prevent a weak signal in one channel from washing out the other.

Late fusion systems commonly combine unimodal decisions using mechanisms such as averaging, 
voting schemes, learned gating, or meta-classifiers~\citep{Atrey2010MultimodalFusion,Baltrusaitis2019MultimodalSurvey}. 
However, rather than introducing a complex secondary model, we employ a straightforward metadata-based dispatcher that dynamically routes the score based on the patch size. 
Commits with small unified diffs are routed to the text channel because patch-level code embeddings carry limited stylometric signal at low change volume. 
Intermediate diffs use an equal-weight blend of the text and code channels, while large diffs are routed to the code channel because they provide richer behavioral fingerprints.

To combine these streams, a two-stage dispatcher first selects the score space for each author. Authors with sufficient prior commit history use per-author percentile ranks computed over previously observed commits, following the streaming anomaly-detection setting~\citep{Chandola2009AnomalySurvey,salehi2018survey}. Sparse histories use raw cosine distance, since percentile estimates are unreliable with few priors. Given the selected scores, the fused anomaly score $s_t$ for commit $c_t$ is determined from the unified diff size $\ell_t$, measured in characters, and predefined lower and upper thresholds $T_{\mathrm{lo}}$ and $T_{\mathrm{hi}}$:
\begin{equation}
s_t =
\begin{cases}
s_t^{\mathrm{text}} & \ell_t < T_{\mathrm{lo}}, \\[4pt]
\frac{1}{2}\left(s_t^{\mathrm{text}} + s_t^{\mathrm{code}}\right) & T_{\mathrm{lo}} \le \ell_t < T_{\mathrm{hi}}, \\[4pt]
s_t^{\mathrm{code}} & \ell_t \ge T_{\mathrm{hi}}.
\end{cases}
\end{equation}
The thresholds $T_{\mathrm{lo}}$ and $T_{\mathrm{hi}}$ are calibrated from modality reliability across commit-size bins on a separate validation split, guided by the patch-size analysis (Section~\ref{sec:loc_auprc}). This patch-size-gated rule fuses single-modality encoder distance streams at score time and complements early-fused encoders trained on concatenated message and diff sequences.

\OceanFigBegin
    \includegraphics[width=\textwidth]{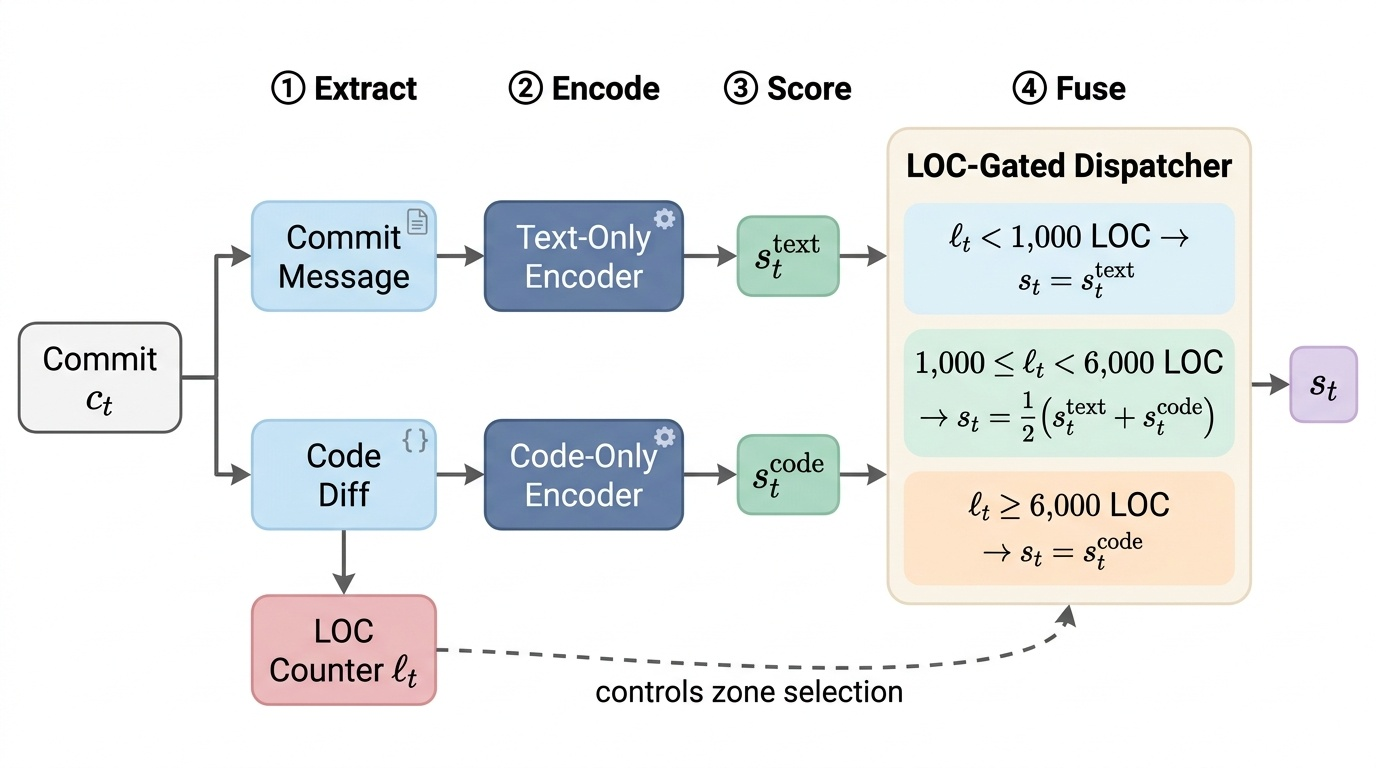}
    \caption{Patch-size-gated fusion dispatcher.}
    \label{fig:arch_c_loc_gate}
\OceanFigEnd

\subsection{Evaluation}
\label{sec:detection_framework}

\textbf{Authorship verification task.} 
To evaluate how well the model distinguishes between commits with the same author,  we construct a balanced (50/50) set of positive and negative pairs from the author-disjoint test split.  
We assess overall discriminative power using the threshold-free Area Under the Receiver Operating Characteristic (ROC AUC)~\citep{Fawcett2006AUC}. 
For operational scenarios, a discrete decision threshold $\tau$ is required to classify pairs. 
We calibrate this threshold on the validation set by selecting the value that maximizes Youden's $J$ index~\citep{Youden1950,Fluss2005}. 
We then apply this calibrated threshold to the test set to compute Accuracy and F1-score. 
We report uncertainty using 95\% confidence intervals derived from a cluster bootstrap over anchor authors ($B=1{,}000$ resamples)~\citep{Efron1994Bootstrap}.

\textbf{Anomaly detection task.} To evaluate how well the model detects anomalous commits in a continuous stream, we apply a chronological sliding window. 
Because anomaly detection is imbalanced, we report AUPRC~\citep{Davis2006AUPRC} for threshold-free ranking and TPR at a fixed 1\% FPR as a low-alert operating point.

Prior work cautions that global metrics can understate the operational false-positive burden in security monitoring~\citep{Sommer2010OutsideWorld,Alahmadi2022SOC,Axelsson2000BaseRate}. 
To directly quantify analyst alert fatigue, we measure the number of benign commits an analyst must review before finding a malicious payload using rank-based metrics defined in this work. 
The specific metric depends on the evaluation scenario. 
For experiments across multiple repositories, we report the Median and Maximum False Positives per true positive (Med.\ FP/pos and Max FP). 
Conversely, for single-repository incident case studies with small positive sample sizes, we report the absolute number of false positives before surfacing the first true positive (FP@1st) and before surfacing all true positives (FP@all). 

Consistent with our verification methodology, we estimate uncertainty for the synthetic detector benchmarks using bootstrap resampling ($B=1{,}000$ resamples) and assess statistical significance with permutation tests~\citep{Efron1994Bootstrap,Good2005Permutation}. 
Bootstrap and permutation procedures, detector sensitivity grids, and extended result tables are in Appendix~\ref{app:detector_details}.
These statistical tests are applied exclusively to the synthetic kernel benchmarks and are not applied to the external incident case studies due to their small positive sample sizes.

%% file: experiments.tex
This subsection specifies the corpora, baselines, and evaluation protocols used to instantiate
Section~\ref{sec:methods}.
We first compare stylometric encoders on an author-disjoint Linux kernel benchmark,
then reuse the trained UniXcoder encoders in a streaming anomaly detector under synthetic
author-swap regimes and two external account-compromise incidents (PHP and ForceMemo).

\subsubsection{Corpora}
\label{sec:experiments_datasets}

\paragraph{Linux kernel (2005--2026).}

The primary dataset is the Linux kernel Git repository~\citep{LinuxKernelRepo}.
We extract commits from the master branch authored between April 2005 and June 2026, yielding 398,505 retained commits from 4,474 authors.
The repository contains approximately 2,500 maintainer-defined subsystem domains derived from the kernel \texttt{MAINTAINERS} file by path matching, covering areas as varied as device drivers, memory management, networking, and cryptography.
Table~\ref{tab:dataset_stats} summarizes sample and author counts per split.

\OceanTabBeginCol
    \caption{Linux kernel corpus statistics by split.}
    \label{tab:dataset_stats}
    \begin{tabular}{lrrr}
        \toprule
        \textbf{Metric} & \textbf{Training} & \textbf{Validation} & \textbf{Test} \\
        \midrule
        Samples & 278,027 & 57,399 & 63,079 \\
        Unique Authors & 3,131 & 671 & 672 \\
        Verification Pairs & -- & 13,420 & 13,440 \\
        \bottomrule
    \end{tabular}
\OceanTabEndCol

\paragraph{PHP backdoor incident (2021).}

On 28 March 2021, two malicious commits were pushed to the core PHP repository under forged maintainer identities, one attributed to Rasmus Lerdorf and one to Nikita Popov~\citep{Popov2021PHPIncident,PHPWatch2021}.
Disguised as minor typo fixes and pushed via compromised credentials, the commits inserted a remote-code-execution backdoor that maintainers intercepted before any official release.
For this external check, we rank commits from the 29 current members of the \texttt{php} GitHub organization who appear in the scored corpus.
The resulting queue contains 26{,}680 commits, including exactly two malicious, misattributed commits as target anomalies.

\paragraph{ForceMemo/GlassWorm (2026).}

The ForceMemo campaign, enabled by GlassWorm credential theft, was a coordinated attack on hundreds of open-source Python-based repositories on GitHub in which adversaries compromised contributor accounts and amended existing legitimate commits, keeping the victim's identity and author date but replacing the code with a backdoored payload and the message with an attacker-controlled disguise (typically short and generic, such as ``Update README.md''), then force-pushed the result into hundreds of repositories~\citep{ForceMemoGlassWorm2026}.
We analyze 40 affected repositories with 5{,}906 commits total.
They contain 62 known malicious, misattributed commits, hereafter \emph{spoofs}, of which 28 remain scoreable after excluding authors who fall below the minimum history threshold after filtering.

\subsubsection{Compared Models}
\label{sec:experiments_models}

We compare three stylometric model families that reflect the approaches reviewed in Sections~\ref{sec:classical-authorship-verification} and~\ref{sec:pretrained-contrastive}.
The first family represents classical lexical stylometry using TF--IDF and character $n$-gram features~\citep{Kestemont2019PAN}, following the FNN lexical baselines of OCEAN~\citep{machtle2024ocean}.
The second uses pre-trained transformer encoders, including UniXcoder~\citep{guo2022unixcoder} and CodeSage-small~\citep{CodeSage2024}\footnote{HuggingFace checkpoint \texttt{codesage-small-v2}.}.
We evaluate both families under the code-only, text-only, and cross-modal input settings of Section~\ref{sec:data_preparation}.
The third adapts SCS-GAN as a neural stylometric baseline~\citep{ou2023scsgan}, and is evaluated in code-only mode.
Architecture and implementation details are in Appendices~\ref{app:fnn_details}, \ref{app:scsgan_details}, and~\ref{app:implementation_details}.

\subsubsection{Authorship Verification Protocol}
\label{sec:rq1_protocol}

We train the compared models on the kernel training split following the representation learning methodology of Section~\ref{sec:training}.
We then apply the verification protocol of Section~\ref{sec:detection_framework} to the Linux kernel validation and test splits (Table~\ref{tab:dataset_stats}).
For each held-out author, we sample exactly ten same-author and ten different-author pairs, yielding twenty pairs per author. 
This fixed sampling budget instantiates the balanced evaluation described in Section~\ref{sec:detection_framework}.
Full hyperparameter configurations and detailed verification metric tables (ROC AUC and AUPRC with confidence intervals) are deferred to Appendices~\ref{app:implementation_details} and~\ref{app:verification_details}.

\subsubsection{Anomaly Detection Task}
\label{sec:rq2_detector_setup}

For the anomaly detection task, no additional training is performed. We reuse the UniXcoder encoders fine-tuned on the 
kernel authorship verification task, including the early-fusion cross-modal encoder and the code-only and text-only single-modality encoders. These embeddings are scored with the chronological detector defined in Section~\ref{sec:detection_framework}.

\paragraph{Synthetic Author Swap.}
To evaluate baseline detection capabilities across modalities, we assign random wrong-author labels to 10\% of commits on the test partition (Section~\ref{sec:detection_framework}), using a 10\% anomaly prevalence as in the controlled
benchmark construction of \citet{Emmott2013AnomalyBenchmarks}.
This dense injection rate also yields a prevalence high enough for stable AUPRC comparisons across encoders and fusion rules~\citep{Davis2006AUPRC}.
This experiment compares single-modality encoders, late-fusion equal-weight blending, and the early-fusion cross-modal encoder using AUPRC and TPR at a strict 1\% FPR.

\paragraph{Patch-Size Analysis.}
To investigate modality reliability as a function of commit size, we reuse the synthetic author swap setting and group commits by diff-character length.
We compare code-only and text-only encoder AUPRC across patch-size bins to empirically ground the thresholds used for patch-size-gated routing.

\paragraph{Patch-Size-Gated Simulation.}
\label{sec:kernel_gated_sim}
To evaluate the operational efficiency of the late-fusion patch-size-gated approach (Section~\ref{sec:score_fusion}), we simulate a sparse short-payload attack by injecting exactly one wrong-author commit for each monitored high-history author (top 20 by commit count, each with $>1{,}000$ commits) on the test partition, restricting positives to diffs under 1{,}000 characters.
This \emph{small-commit} setting matches the text-dominant regime identified by the patch-size analysis.
We quantify review burden using Median and Maximum FP per positive, comparing the patch-size-gated dispatcher against single-channel and equal-mix rules on this maintainer-scoped queue.
The patch-size-gated dispatcher uses thresholds calibrated on a separate validation split ($T_{\mathrm{lo}} = 1{,}000$ and $T_{\mathrm{hi}} = 6{,}000$ characters), guided by the patch-size analysis (Section~\ref{sec:loc_auprc}).
An unrestricted-injection sensitivity check, where the injected commit may have any size, is reported in Appendix~\ref{app:kernel_fusion}.

\paragraph{PHP Incident.}
For the PHP corpus (Section~\ref{sec:experiments_datasets}), we apply the patch-size-gated dispatcher configured using kernel validation data, as evaluated in Section~\ref{sec:kernel_gated_sim}, without retraining on the incident corpus.
Because the PHP incident involved a server-level compromise of core maintainers' credentials, the threat model requires organization-wide screening of high-trust contributors. We therefore rank PHP commits in a single maintainer-scoped chronological queue.
Authors with fewer than 50 prior commits are excluded, and we reuse the kernel sliding-window size $W=20$ without retuning on the incident corpus.
We report false positives before surfacing the first and all forged commits (FP@1st and FP@all).

\paragraph{ForceMemo Incident.}
For the ForceMemo corpus (Section~\ref{sec:experiments_datasets}), we similarly apply the patch-size-gated dispatcher configured using kernel validation data, as evaluated in Section~\ref{sec:kernel_gated_sim}, without retraining on the incident corpus.
Because many compromised contributors have sparse histories, the sliding-window length is reduced for scoreability rather than retuned for detection performance (Appendix~\ref{app:fm_case_study_details}).
Because the GlassWorm campaign exploited multiple repositories, we rank ForceMemo spoofs within each affected repository's commit history.
Review burden is summarized with Med.\ FP/pos and Max FP across the 28 per-repository rankings.

Supplementary detector settings, fusion grids, full rule tables, and extended case-study scoring details are in Appendices~\ref{app:detector_details}, \ref{app:kernel_fusion}, \ref{app:php_case_study_details}, and~\ref{app:fm_case_study_details}.

%% file: results.tex
\hyphenation{UniX-coder Code-Sage}

This section reports the empirical outcomes of the experiments described in Section~\ref{sec:experiments}.
We first report authorship verification on held-out kernel pairs, then anomaly detection under synthetic author swap, patch-size analysis, patch-size-gated simulation, and external ranking checks on the PHP and ForceMemo incident corpora.

\subsection{Authorship Verification Task}
\label{sec:rq1_verification_results}

Table~\ref{tab:fusion_results} summarizes open-world verification on held-out test pairs.

\OceanTabBeginCol
    \OceanTabCaption{Kernel verification ROC AUC [95\% CI] on held-out test pairs (higher is better)}
    \label{tab:fusion_results}
    \OceanResizeTable{
    \begin{tabular}{lccc}
        \toprule
        \textbf{Model Name} & \textbf{Code-Only} & \textbf{Text-Only} & \textbf{Cross-Modal} \\
        \midrule
        SCS-GAN (adapted baseline) & 0.675 [0.663, 0.687] & -- & -- \\
        FNN TF-IDF & 0.719 [0.707, 0.731] & 0.750 [0.737, 0.763] & 0.702 [0.691, 0.714] \\
        FNN N-gram & 0.834 [0.823, 0.845] & 0.835 [0.823, 0.847] & 0.828 [0.817, 0.839] \\
        CodeSage-small-v2 & 0.905 [0.896, 0.914] & 0.915 [0.905, 0.924] & 0.927 [0.919, 0.934] \\
        \textbf{UniXcoder} & \textbf{0.907 [0.898, 0.916]} & \textbf{0.931 [0.923, 0.938]} & \textbf{0.932 [0.924, 0.939]} \\
        \bottomrule
    \end{tabular}
    }
\OceanTabEndCol

Pre-trained encoders outperformed lexical baselines and the adapted SCS-GAN baseline on every reported modality.
UniXcoder recorded the highest scores, with text-only and cross-modal nearly tied at $0.93$ ROC AUC (exact values $0.9308$ and $0.9322$).
Relative to the weakest lexical cross-modal entry (FNN TF-IDF, $0.702$), this is a gain of $0.230$ ROC AUC.
Relative to FNN N-gram cross-modal ($0.828$) and CodeSage-small-v2 cross-modal ($0.927$), the UniXcoder gains are $0.105$ and $0.006$ ROC AUC.
Full threshold-free metrics with 95\% CIs (ROC AUC, AUPRC) appear in Appendix~\ref{app:verification_details} (Tables~\ref{tab:code_results_appendix}--\ref{tab:fusion_results_appendix}).
Appendix~\ref{app:implementation_details} reports an equal-truncation sensitivity check in which cross-modal exceeded text-only by $0.0065$ ROC AUC ($0.9195$ vs.\ $0.9130$).

\subsection{Anomaly Detection Task}
\label{sec:rq2_anomaly_detection}

\paragraph{Synthetic Author Swap.}
Table~\ref{tab:kernel_global_swap} reports UniXcoder anomaly detection under 10\% synthetic author swap on the kernel test partition.
Rows compare separately trained single-modality encoders and the early-fusion cross-modal encoder (Section~\ref{sec:detection_framework}).
TPR@1\% FPR is the fraction of injected commits detected when the false-alarm rate on benign commits is capped at 1\%.

\OceanTabBeginCol
    \OceanTabCaption{Synthetic author swap: UniXcoder detection (higher is better).}
    \label{tab:kernel_global_swap}
    \begin{tabular}{@{}llcc@{}}
        \toprule
        \textbf{Scoring} & \textbf{Channel} & \textbf{AUPRC [95\% CI]} & \textbf{TPR@1\% FPR} \\
        \midrule
        Encoder  & Code-only  & 0.916 [0.904, 0.927] & 0.627 \\
        Encoder  & Text-only  & 0.946 [0.937, 0.953] & 0.772 \\
        \textbf{Early-fusion} & \textbf{Cross-modal} & \textbf{0.939 [0.929, 0.947]} & \textbf{0.737} \\
        \bottomrule
    \end{tabular}
\OceanTabEndCol

Text-only and early-fusion cross-modal training achieved comparable, highest ranking quality (AUPRC $0.946$ and $0.939$), both exceeding code-only ($0.916$) and remaining far above chance.
Under the fixed low-false-positive operating point (1\% FPR), text-only again recorded the highest sensitivity ($77.2\%$ TPR), above early-fusion cross-modal ($73.7\%$) and code-only ($62.7\%$), so the text-only channel leads on both ranking quality and low-FPR sensitivity while cross-modal is comparable on AUPRC but trails text-only at this operating point.
Appendix~\ref{app:kernel_fusion} reports the full fusion ablation.
Appendix Table~\ref{tab:kernel_detection} places the same UniXcoder cross-modal result against lexical detectors under the identical protocol (FNN TF--IDF cross-modal AUPRC $0.217$, FNN N-gram cross-modal $0.628$).

\paragraph{Patch-Size Analysis.}
\label{sec:loc_auprc}
Figure~\ref{fig:loc_auprc} reports anomaly detection performance across different patch sizes.

\OceanFigBegin
    \includegraphics[width=0.82\textwidth]{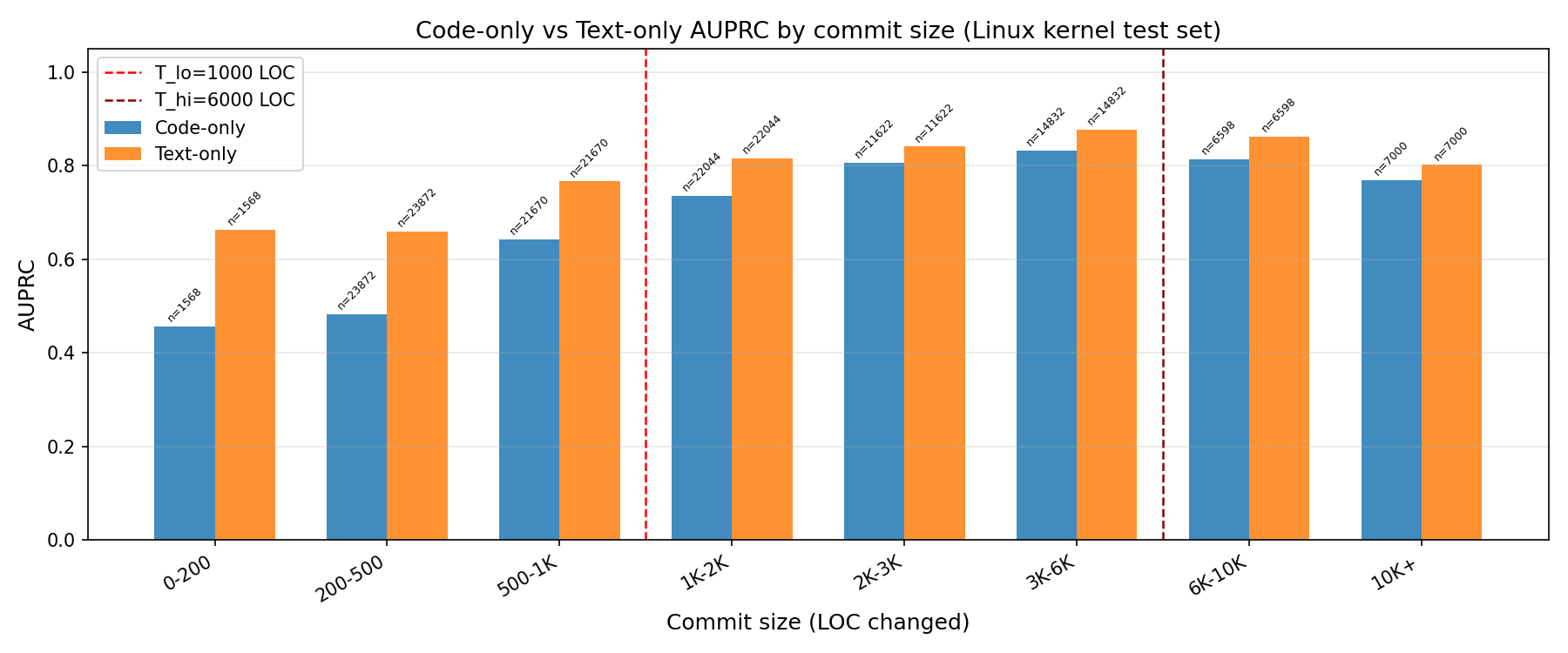}
    \OceanFigCaption{AUPRC by patch-size bin (code-only vs.\ text-only). Higher is better.}
    \label{fig:loc_auprc}
\OceanFigEnd

For small commits (under 1{,}000 characters), the text-only channel outperformed the code-only channel by up to $0.206$ AUPRC.
For larger commits (over 1{,}000 characters), this gap narrowed to $0.079$ AUPRC or less.

\paragraph{Patch-Size-Gated Simulation.}
Table~\ref{tab:kernel_results_summary} reports review burden on the top-20 maintainer queue ($\approx 60{,}571$ commits) under one-per-author small-commit injection (Section~\ref{sec:rq2_detector_setup}).
Med.\ FP/pos is the median number of benign commits ranked above a positive, and Max FP is the worst-case count across positives (Section~\ref{sec:detection_framework}).
Lower values indicate less review work before the typical or worst injected commit.

\OceanTabBeginCol
    \OceanTabCaption{Patch-size-gated simulation: review burden under one-per-author small-commit injection (diffs under 1{,}000 characters). Med.\ FP/pos is reported with a 95\% bootstrap confidence interval over pooled injected positives. Lower is better.}
    \label{tab:kernel_results_summary}
    \begin{tabular}{@{}lcr@{}}
        \toprule
        \textbf{Rule} & \textbf{Med.\ FP/pos [95\% CI]} & \textbf{Max FP} \\
        \midrule
        Text-only raw          & 81 [30, 145] & 3{,}008 \\
        Code-only raw          & 886 [116, 1{,}024] &  4{,}720 \\
        Equal mix raw          & 65 [14, 113] &  2{,}703 \\
        \midrule
        Patch-size-gated dispatcher   & 50 [28, 225] & 1{,}843 \\
        \bottomrule
    \end{tabular}
\OceanTabEndCol

On the primary metric (Med.\ FP/pos), the text-only channel and the two fusion rules achieved comparable review burden (median $81$, $65$, and $50$ false positives per positive for text-only, equal-mix, and the dispatcher, about $0.08$--$0.13\%$ of the $\approx 60{,}571$-commit queue), while the code-only channel was far worse ($886$).
Their marginal 95\% bootstrap intervals overlap (Table~\ref{tab:kernel_results_summary}), but a paired bootstrap over the same injected positives shows the code-only channel is significantly worse than each of the other three rules (median per-positive FP reduction of $405$ to $533$, with 95\% intervals excluding zero), while text-only, equal-mix, and the dispatcher are statistically indistinguishable from one another. This small simulation (four to eleven positives per seed) therefore separates only the code-only channel.
The dispatcher recorded the lowest worst-case burden (Max FP $1{,}843$).
This pattern matches the text-dominant short-payload regime identified by the patch-size analysis (Section~\ref{sec:loc_auprc}): for small commits the text channel alone is competitive, and adding the code channel through fusion neither clearly helps nor hurts.

\paragraph{PHP Incident Evaluation.}
\label{sec:incident_validation_php}
Table~\ref{tab:php_results_summary} ranks the PHP maintainer queue of 26{,}680 commits, which contains two forged commits (Section~\ref{sec:experiments_datasets}).
Lower values indicate fewer false alarms before finding the forged commits.

\OceanTabBeginCol
    \OceanTabCaption{PHP incident: the two forged commits ranked in a single org-wide queue (lower is better). FP@1st / FP@all count benign commits ranked above the first / all forged commits.}
    \label{tab:php_results_summary}
    \begin{tabular}{@{}lrr@{}}
        \toprule
        \textbf{Rule} & \textbf{FP@1st} & \textbf{FP@all} \\
        \midrule
        Text-only raw          & 309 & 605 \\
        Code-only raw          & 11{,}700 & 13{,}622 \\
        Equal mix raw          & 1{,}186 & 2{,}305 \\
        \textbf{Patch-size-gated dispatcher} & \textbf{248} & \textbf{274} \\
        \bottomrule
    \end{tabular}
\OceanTabEndCol

The patch-size-gated dispatcher, using per-author percentile scores computed over prior commits only, yielded the lowest false-positive counts: 248 before the first forgery and 274 before all (about $1.03\%$ of the 26{,}680-commit queue, Table~\ref{tab:php_results_summary}).
This is 331 fewer false positives than the raw text-only baseline, which required 605.
The equal-mix rule required $2{,}305$ false positives before all forgeries, and the code-only rule required $11{,}700$ before the first.
Because this incident contains only two positives, these numbers represent a specific case study rather than a general detection rate.

\paragraph{ForceMemo Incident Evaluation.}
\label{sec:incident_validation_fm}
Table~\ref{tab:fm_results_summary} summarizes this review burden across all 28 repositories using the median and maximum false positives per spoof.

\OceanTabBeginCol
    \OceanTabCaption{ForceMemo incident: 28 scoreable spoofs ranked within each repository (lower is better).}
    \label{tab:fm_results_summary}
    \begin{tabular}{@{}lrr@{}}
        \toprule
        \textbf{Rule} & \textbf{Med.\ FP/pos} & \textbf{Max FP} \\
        \midrule
        Text-only raw           & 6 (21.4\%) & 262 (87.3\%) \\
        Code-only raw           & 1 (6.1\%) & 110 (36.7\%) \\
        Equal mix raw           & 3 (7.7\%) & 243 (81.0\%) \\
        \textbf{Patch-size-gated dispatcher} & \textbf{1 (0.8\%)} & \textbf{126 (42.0\%)} \\
        \bottomrule
    \end{tabular}
\OceanTabEndCol

On the primary metric, the patch-size-gated dispatcher and the code-only rule tied at a median of 1 false positive per spoof (Table~\ref{tab:fm_results_summary}).
This means that for the typical repository in this campaign, an analyst would only need to review one benign commit before finding the attacker's payload.
This outperformed both the text-only rule (median 6) and the equal-mix rule (median 3).
Expressed as a share of each repository's scored queue, the dispatcher's median burden is $0.8\%$, compared with $6.1\%$ for code-only, $7.7\%$ for equal mix, and $21.4\%$ for text-only.
On the secondary worst-case metric, code-only was slightly better ($110$ Max FP) than the dispatcher ($126$), while text-only was worst ($262$).
All four rules' worst cases occur in the same repository, \seqsplit{BrianElionDev/BuyBot} (324 commits total, 300 in its scored ranking queue).
There the dispatcher and code-only rule respectively require reviewing $42.0\%$ and $36.7\%$ of the queue before reaching the spoof, versus the campaign-wide median of one commit.
Appendix~\ref{app:fm_case_study_details} reports a residual-case analysis of the four repositories with the heaviest review burden.

\FloatBarrier

%% file: discussion.tex
This study demonstrates that patch-level authorship verification and anomaly detection are feasible in a large open-world repository setting. Both the code patch and the accompanying commit message carry a strong author-specific signal, with cross-modal and text-only representations nearly tied at the top of the verification ranking. Furthermore, when deployed as a streaming anomaly detector, these representations place forged commits high in retrospective audit rankings, and a dynamic routing mechanism can adapt to varying patch sizes to manage the typical false-positive burden.

Addressing RQ1 (patch-level authorship signal), the results reveal a clear modality hierarchy: text-only representations consistently outperform code-only representations on average, and cross-modal UniXcoder is marginally highest on the reported point estimate (0.9322 vs.\ 0.9308 for text-only, Table~\ref{tab:fusion_results}). This indicates that natural language phrasing, intent description, and formatting in commit messages provide a denser stylistic fingerprint than structural code choices alone.

The model-comparison experiment indicates that encoder choice matters less than modality alignment with the patch-level task. CodeSage-small-v2 is a newer code representation model, yet UniXcoder performs better across the reported modalities in this benchmark (Table~\ref{tab:fusion_results}).
This result suggests that patch-level stylometry may reward representations that preserve cross-modal code-natural-language cues and surface regularities rather than semantic code understanding alone.

While the text signal provides the strongest empirical verification performance, from a threat-model perspective, it is inherently easier for an adversary to impersonate by copying public commit logs. Conversely, the code signal imposes a higher mimicry burden because forged code must remain functional and fit project APIs. The small performance gap (+0.0014 AUC) between the cross-modal model and the text-only encoder reflects this trade-off, showing that while the code signal adds limited average benefit in a benign setting, it may still provide a harder-to-mimic channel than commit messages, though this was not evaluated under targeted mimicry.

Addressing RQ2 (operational misuse detection), the anomaly-detection experiments show how the learned representations behave when converted into an anomaly detector. The synthetic author swap experiment establishes separability under dense label mismatch, with the text-only and cross-modal encoders achieving the highest AUPRC (0.946 and 0.939, Table~\ref{tab:kernel_global_swap}). Under the fixed 1\% FPR operating point, text-only also records the highest TPR, so it leads on both ranking and sensitivity, while cross-modal trails at that operating point. The patch-size analysis (Figure~\ref{fig:loc_auprc}) explains why a single modality is insufficient across all commits: text-only signal dominates short patches, while the text--code gap narrows for larger payloads.

The patch-size-gated simulation applies the patch-size observation in a realistic short-payload queue. Under one-per-author small-commit injection, the text-only channel, the equal-mix blend, and the patch-size-gated dispatcher reduce the median false-positive burden to comparable levels (81, 65, and 50 commits, roughly 0.08--0.13\% of the $\approx 60{,}571$-commit queue, Table~\ref{tab:kernel_results_summary}), far below the code-only channel. Their confidence intervals overlap within the noise of this small simulation, which separates only code-only from the rest, so we read it as consistent with the text-dominant short-payload regime rather than as ranking the fusion rules. When the injected commit is not restricted by size, equal-mix raw scoring is clearly strongest (Appendix~\ref{app:kernel_fusion}), so the dispatcher should be read as a short-payload routing rule rather than a universal default.

The PHP incident evaluation provides a cross-repository and cross-language check for short forged commits. Both forged commits fell below the 1{,}000-character routing threshold, so the patch-size-gated dispatcher routed them through the text-dominant channel. The dispatcher surfaced both malicious commits in the top 1.03\% of the retrospective audit queue (Table~\ref{tab:php_results_summary}). This result provides preliminary anecdotal support that the kernel-trained detector is consistent with transfer to a different ecosystem without retraining, but with only two positives, it cannot establish robust generalization.

The ForceMemo incident evaluation tests a different regime, where spoofed commits are larger and payload-heavy. In this setting, the signal shifts toward the code channel. All scoreable spoofs exceeded the upper patch-size gate, causing the dispatcher to correctly route them to the code channel without text fusion. The dispatcher achieved a median of 1 FP/pos, tying the code-only encoder, though it did not outperform code-only on worst-case ranking (max FP, Table~\ref{tab:fm_results_summary}). This complements the PHP result by showing that the routing strategy can adapt when the discriminative signal moves from commit-message behavior to code-patch behavior. Together, the two incident evaluations indicate that the detector's ranking behavior is consistent with the modality mechanism observed in the kernel experiments.

The per-repository breakdown of the ForceMemo result also shows where the stylometric layer is strong and where it is not. The four repositories with the heaviest residual review burden are associated with impersonated accounts whose histories mix bulk binary or data drops with repetitive large edits, so the spoof sits nearer the middle of that account's own distance distribution (Appendix~\ref{app:fm_case_study_details}). A stylometric verifier can only flag deviation from an established norm, so accounts without a stable baseline receive weaker protection from this layer. Because that condition depends only on pre-attack history, such accounts can in principle be identified in advance and prioritized for complementary content-level screening. This study does not evaluate that screening layer, but the GlassWorm payloads' large encoded blobs make content scanning a natural complement. Read this way, the residual cases argue for layering authorship verification with content checks rather than for discarding either mechanism.

This open-world patch-level capability extends classical code stylometry~\citep{Caliskan2015}, which historically struggled with short, incomplete fragments~\citep{Dauber2019GitBlameWho}. Topic confounding in natural-language authorship verification provides a related analogy for shared vocabulary across authors~\citep{seroussi2014topicmodels}. Our contrastive approach aligns methodologically with frameworks like OCEAN \citep{machtle2024ocean} and CLAVE \citep{alvarez2025clave}, which frame open-world authorship verification with cosine-similarity embeddings, and the adversarial SCS-GAN \citep{ou2023scsgan}. However, this study differs from these precedents in evaluation scale and sample granularity. Where OCEAN targets supply-chain injection detection on compiled functions and CLAVE reports high AUROC on Python files, we demonstrate that pre-trained representations can extract reliable authorial signatures from unified version-control patches. Furthermore, the strong performance of text-only message stylometry recontextualizes the findings of \citet{Dauber2019GitBlameWho}, who showed that code-fragment attribution is exceptionally difficult. By treating the commit as a cross-modal artifact rather than just a code fragment, the verification task becomes tractable.

The evaluation combines controlled synthetic author swaps with retrospective analyses of two real-world account-compromise incidents, PHP and ForceMemo. These external cases demonstrate relevance to observed compromises, but they do not evaluate adaptive attackers who deliberately imitate a maintainer's style or gradually alter the historical baseline. Sparse author histories also remain a practical limitation because reliable anomaly scoring requires sufficient prior commits.

A further limitation is the growing use of large language models in software development. Because the evaluated corpora span recent years, some commits may include AI-assisted code or messages. We neither detected nor filtered such content, so its effect on author representations is unknown. This remains an open question for future evaluation rather than a claim supported by the present results.

%% file: conclusions.tex
The global software supply chain relies on signatures, access controls, and repository permissions that attest to who authorized a change, yet these mechanisms provide limited behavioral signal when credentials or maintainer accounts are misused. This study examined whether patch-level authorship verification can help close that identity gap. By treating code diffs and commit messages as complementary behavioral artifacts, the results show that supervised contrastive representations can verify authorship on held-out kernel commits and rank forged commits in retrospective supply-chain incident queues.

The central contribution is therefore twofold. First, the verification results show that patch-level commits contain measurable author-specific signal at repository scale, with UniXcoder text-only and cross-modal variants nearly tied at the top of the held-out ranking. Second, the PHP and ForceMemo incident evaluations show that this signal is operationally meaningful beyond synthetic benchmarks.

The PHP case represents short forged commits where commit-message behavior is most informative, while ForceMemo represents large payload-heavy spoofs where code-patch behavior carries the stronger signal. Together, these cases support the broader claim that behavioral identity checks can complement provenance controls across different attack shapes.

The patch-size-gated dispatcher should be interpreted as a simple demonstration of late-fusion value rather than as a final deployment architecture. Its purpose in this study was to show that routing between text and code channels can reduce review burden when the dominant stylometric signal changes with patch size. This finding is useful because it makes the modality trade-off explicit, but the rule itself is intentionally simple. Future systems should evaluate learned or probabilistic fusion mechanisms that incorporate patch size, author history, repository context, uncertainty, and attack-specific priors.

These findings imply that patch-level behavioral verification is best viewed as a calibrated triage layer, not a replacement for signatures, access controls, review, or provenance mechanisms. In a continuous integration setting, the detector should produce ranked queues for analyst review rather than autonomous block decisions. This framing is especially important because real impersonation events are rare, author histories can be highly variable, and sparse-history identities may be difficult or impossible to score. The ForceMemo evaluation illustrates the related constraint that accounts without a stable pre-attack baseline receive weaker protection from this layer.

Several limitations remain. The synthetic author-swap experiments model \emph{na\"ive impersonation}, where labels are inconsistent with code and message content, and therefore provide an easiest-case diagnostic rather than a complete adversarial evaluation. They do not model targeted mimicry, gradual account takeover, poisoning of author baselines, or deliberate manipulation of patch size to influence routing. Topic controls also remain imperfect, since subsystem matching is only a coarse proxy for topical overlap, and although we resolve author identity by linking shared emails and name spellings, rare reverse collisions (distinct contributors sharing an address through applied-patch metadata) cannot be fully excluded.

Future work should therefore move from simple late fusion and \emph{na\"ive} swaps toward stronger adversarial and deployment evaluations. Important directions include domain-matched impersonation, targeted stylistic mimicry, learned modality fusion, adaptive and time-aware author baselines, sparse-history fallback strategies, and integration with metadata such as review behavior, signing status, repository role, and social trust signals. Future work should also examine stylometric defenses under AI-assisted development, including whether different models leave distinct stylistic signatures and whether human prompting still shapes the final patch and commit message.

Ultimately, patch-level behavioral identity checks offer a promising path toward trust-verifiable software maintenance when they are deployed as calibrated signals within a broader defense-in-depth workflow.

\paragraph{Code availability.}
The training, evaluation, and anomaly-detection source code and configurations used in this study will be made publicly available upon publication.

%% file: appendix_a.tex
\section{Additional Experimental Details}

\subsection{Baseline Architecture Details}
\label{app:fnn_details}

For the statistical baselines described in Section~\ref{sec:experiments_models}, we follow the feedforward lexical baselines of OCEAN~\citep{machtle2024ocean}: a shared MLP on top of TF--IDF character features. \textbf{FNN TF--IDF} used character unigrams with a capped vocabulary of 1{,}000 features, while \textbf{FNN char $n$-gram} used character $n$-grams with \(n\in\{1,2,3\}\) and the same feature cap. Character $n$-gram features are a standard lexical baseline in authorship evaluation~\citep{Kestemont2019PAN}. In both cases, the resulting feature vector was mapped through an MLP with dimensions $|V|\rightarrow 64 \rightarrow 64 \rightarrow 768$ before the shared projection head used for verification.

\subsection{SCS-GAN Adaptation Details}
\label{app:scsgan_details}

The published SCS-GAN~\citep{ou2023scsgan}\footnote{Implementation source: \texttt{https://github.com/L1NNA/SourceCodeAuthorshipAnalysis} (branch \texttt{SCS-Gan}, path \texttt{src/SCS-Gan}).} targets authorship verification on whole Google Code Jam (GCJ) source files. Adapting it to our open-world binary verification setting on patch-level commits changed the input domain and evaluation modality while retaining the original adversarial architecture:

\begin{enumerate}
    \item \textbf{Input format.} The original model operates on whole source files. We replaced the input with our linearized code diff (Section~\ref{sec:methods}), tokenized with a SentencePiece subword vocabulary (size 30{,}000) retrained on the kernel corpus and truncated to a fixed length. Consistent with the code-only setting, the commit message is not included in the SCS-GAN input.

    \item \textbf{Architecture and training.} We retained SCS-GAN's original adversarial design, a discriminator with multi-head stylistic attention trained against a generator, and trained it on our patch corpus with the original GAN objective. 
    The verification score is the cosine similarity between the encoder outputs, matching the evaluation protocol for the other baselines.

    \item \textbf{Modality.} SCS-GAN was evaluated only in code-only mode, as the original model's multi-head attention architecture for isolating stylistic tokens was designed for source-code inputs. Extending it to commit messages would require revalidating the attention-head design choices, which is outside the scope of this comparison.
\end{enumerate}

Code-only verification scores for the adapted baseline are reported with the other models in Table~\ref{tab:fusion_results} and Table~\ref{tab:code_results_appendix}.

\subsection{Additional Implementation Details}
\label{app:implementation_details}

\textbf{Author and commit retention.}
Authors must have at least ten qualifying commits under the training diff-length bounds applied during corpus construction.
Commits outside those bounds are excluded from training and evaluation splits.

Key hyperparameters and implementation details are:
\begin{itemize}
    \item \textbf{Optimizer:} AdamW (weight decay $3\times10^{-4}$).
    \item \textbf{Learning rate:} $2\times10^{-5}$.
    \item \textbf{Batch size:} 12 (training. Embedding and inference batching in the detector defaults to 8). Training uses the unit-based loader in Section~\ref{sec:training} with $u=2$ commits per author unit, so each step forms $K=B/u=6$ disjoint within-author pairs.
    \item \textbf{Epochs:} 5.
    \item \textbf{SupCon temperature:} $\tau_{\mathrm{con}} = 0.08$.
    \item \textbf{Projection head:} a three-layer MLP $768 \rightarrow 512 \rightarrow 512 \rightarrow 256$ with ReLU activations and dropout ($0.1$) between layers. The 256-dimensional output is $\ell_2$ normalized before cosine similarity.
    \item \textbf{Hardware:} NVIDIA RTX 6000 Ada with 48 GB GDDR6 VRAM.
\end{itemize}

\subsection{Detailed Verification Results}
\label{app:verification_details}

This appendix reports the full per-modality verification tables, including bootstrap confidence intervals for each model. Table~\ref{tab:code_results_appendix} reports code-only results, Table~\ref{tab:text_results_appendix} reports text-only results, and Table~\ref{tab:fusion_results_appendix} reports cross-modal results. These tables complement the consolidated ROC AUC comparison in the main Results section (Table~\ref{tab:fusion_results}). AUPRC is included as a supplementary metric for completeness, though it is not emphasized in the main text because the verification pairs are balanced.

\textbf{Token-budget sensitivity.}
In the production configuration, the cross-modal input truncates the commit message to 200 characters before concatenation with the code diff, while the text-only encoder uses a 3{,}000-character message limit. We retrained the UniXcoder text-only and cross-modal variants with the same 3{,}000-character message limit before tokenizer truncation. Under this equal-truncation configuration, text-only obtains ROC AUC 0.9130 and cross-modal obtains ROC AUC 0.9195. Uncertainty intervals were not computed for this auxiliary run.

\OceanTabBegin
    \caption{Code-only verification results.}
    \label{tab:code_results_appendix}
    \OceanResizeTable{
    \begin{tabular}{lcc}
        \toprule
        \textbf{Model Name} & \textbf{ROC AUC [95\% CI]} & \textbf{AUPRC [95\% CI]} \\
        \midrule
        UniXcoder                           & \textbf{0.9070 [0.8975--0.9162]} & \textbf{0.9187 [0.9112--0.9259]} \\
        CodeSage-small-v2                   & 0.9047 [0.8955--0.9137] & 0.9155 [0.9075--0.9237] \\
        FNN N-gram                          & 0.8338 [0.8231--0.8447] & 0.8306 [0.8190--0.8428] \\
        FNN TF-IDF                          & 0.7187 [0.7072--0.7311] & 0.7240 [0.7105--0.7374] \\
        SCS-GAN (adapted baseline)          & 0.6752 [0.6634--0.6867] & 0.6608 [0.6482--0.6739] \\
        \bottomrule
    \end{tabular}
    }
\OceanTabEnd

\OceanTabBegin
    \caption{Text-only verification results.}
    \label{tab:text_results_appendix}
    \OceanResizeTable{
    \begin{tabular}{lcc}
        \toprule
        \textbf{Model Name} & \textbf{ROC AUC [95\% CI]} & \textbf{AUPRC [95\% CI]} \\
        \midrule
        UniXcoder                           & \textbf{0.9308 [0.9229--0.9378]} & \textbf{0.9413 [0.9348--0.9478]} \\
        CodeSage-small-v2                   & 0.9150 [0.9054--0.9242] & 0.9290 [0.9216--0.9364] \\
        FNN N-gram                          & 0.8348 [0.8233--0.8467] & 0.8350 [0.8231--0.8455] \\
        FNN TF-IDF                          & 0.7501 [0.7372--0.7628] & 0.7629 [0.7509--0.7750] \\
        \bottomrule
    \end{tabular}
    }
\OceanTabEnd

\OceanTabBegin
    \caption{Cross-modal verification results.}
    \label{tab:fusion_results_appendix}
    \OceanResizeTable{
    \begin{tabular}{lcc}
        \toprule
        \textbf{Model Name} & \textbf{ROC AUC [95\% CI]} & \textbf{AUPRC [95\% CI]} \\
        \midrule
        UniXcoder                           & \textbf{0.9322 [0.9240--0.9393]} & \textbf{0.9427 [0.9361--0.9484]} \\
        CodeSage-small-v2                   & 0.9265 [0.9192--0.9339] & 0.9376 [0.9311--0.9436] \\
        FNN N-gram                          & 0.8276 [0.8174--0.8389] & 0.8210 [0.8082--0.8332] \\
        FNN TF-IDF                          & 0.7022 [0.6906--0.7143] & 0.7108 [0.6983--0.7226] \\
        \bottomrule
    \end{tabular}
    }
\OceanTabEnd

\subsection{Detailed Anomaly Detection Results}
\label{app:detector_details}

This appendix supplements the kernel anomaly-detection protocol defined in Section~\ref{sec:rq2_detector_setup} with ablations, confidence intervals, and full result grids.
Unless otherwise stated, the detector uses window size $W=20$ and cosine distance aggregated by the mean over the previous window.
Threshold-dependent detector metrics use a Youden-calibrated operating point on a held-out fraction of the test stream. Rank-based review-burden metrics do not use this threshold.
Confidence intervals for synthetic detector benchmarks use commit-level bootstrap resampling ($n=1{,}000$), distinct from the author-cluster bootstrap used for verification in Section~\ref{sec:detection_framework}. Hypothesis tests use permutation tests ($n=1{,}000$)~\citep{Good2005Permutation,Efron1994Bootstrap}.

\textbf{Evaluation experiments.} The detector experiments operate on a full-range test split (commits of every size) over the same 672 held-out authors as the verification benchmark, comprising $178{,}932$ commits. This is a superset of the length-restricted verification test partition in Table~\ref{tab:dataset_stats} ($63{,}079$ commits), because the detector scores commits of all sizes rather than the training-length band. The synthetic author swap experiment uses this full-range test split (672 authors); the patch-size analysis uses the corresponding validation split.
The patch-size-gated simulation instead restricts monitoring to the top-20 kernel authors by commit count (each with $>1{,}000$ commits), yielding a maintainer-scoped queue of ${\approx}60{,}571$ commits with a per-commit positive rate of ${\approx}0.01\%$ under one-per-author injection (about four to eleven positives per seed).

Table~\ref{tab:window_sweep} shows AUPRC and ROC AUC for the cross-modal UniXcoder detector across $W \in \{3,5,8,10,15,20\}$ under the 10\% synthetic author swap benchmark. ROC AUC is included here as a diagnostic ablation metric, while AUPRC remains the primary measure of performance under imbalance (Section~\ref{sec:detection_framework}).
AUPRC peaks at $W=10$--$15$ ($0.943$) and reaches $0.939$ at $W=20$. We retain $W=20$ to use the longest evaluated history and reuse it without tuning for the held-out PHP audit. The ForceMemo deviation to $W=5$ is documented in Appendix~\ref{app:fm_case_study_details}.

\OceanTabBegin
    \caption{Detector window-size ablation ($W$).}
    \label{tab:window_sweep}
    \begin{tabular}{ccc}
        \toprule
        $W$ & AUPRC & ROC AUC \\
        \midrule
        3  & 0.894 & 0.958 \\
        5  & 0.924 & 0.980 \\
        8  & 0.939 & 0.989 \\
        10 & 0.943 & 0.990 \\
        15 & 0.943 & 0.991 \\
        \textbf{20} & \textbf{0.939} & \textbf{0.991} \\
        \bottomrule
    \end{tabular}
\OceanTabEnd

\OceanTabBegin
    \caption{Kernel synthetic author swap detection ($W=20$).}
    \label{tab:kernel_detection}
    \OceanResizeTable{
    \begin{tabular}{llcc}
        \toprule
        \textbf{Model} & \textbf{Modality} & \textbf{AUPRC [95\% CI]} & \textbf{TPR@1\% FPR} \\
        \midrule
        UniXcoder & Code  & 0.916 [0.904, 0.927] & 0.627 \\
        UniXcoder & Text  & \textbf{0.946 [0.937, 0.953]} & 0.772 \\
        UniXcoder & Cross-modal &0.939 [0.929, 0.947] & 0.737 \\
        \midrule
        CodeSage-small-v2 & Code  & 0.817 [0.783, 0.851] & 0.672 \\
        CodeSage-small-v2 & Text  & 0.863 [0.837, 0.890] & 0.776 \\
        CodeSage-small-v2 & Cross-modal &0.805 [0.773, 0.837] & 0.658 \\
        \midrule
        FNN TF--IDF & Code  & 0.292 [0.249, 0.339] & 0.128 \\
        FNN TF--IDF & Text  & 0.264 [0.221, 0.311] & 0.128 \\
        FNN TF--IDF & Cross-modal &0.217 [0.169, 0.273] & 0.074 \\
        \midrule
        FNN N-gram & Code  & 0.638 [0.597, 0.681] & 0.401 \\
        FNN N-gram & Text  & 0.648 [0.604, 0.694] & 0.416 \\
        FNN N-gram & Cross-modal &0.628 [0.586, 0.675] & 0.399 \\
        \bottomrule
    \end{tabular}
    }
\OceanTabEnd

\subsection{Kernel Late-Fusion and Routing Ablation}
\label{app:kernel_fusion}
\label{app:kernel_ablation}

We evaluate the equal-mix and patch-size-gated late-fusion strategies defined in Section~\ref{sec:score_fusion}. For UniXcoder, equal-weight late fusion does not exceed the jointly trained cross-modal encoder, although the two results use independent noise draws and are not compared directly.

To separate the contributions of patch-size routing and per-author normalization, we evaluate both noise regimes on UniXcoder using a two-zone gate (code length below $1{,}000$ routes to text, all other commits use the equal mix), with percentile variants computed in prior-only mode.
Under synthetic author swap noise the equal mix records AUPRC $0.915$, compared with $0.873$ for the patch-size-gated pipeline.
For commits under $1{,}000$ characters, the equal mix records AUPRC $0.881$, compared with $0.858$ for text-only and $0.722$ for code-only. Per-author normalization does not yield a useful global ranking under fixed-fraction, per-author injection because each author receives the same relative anomaly prevalence.

\textbf{Unrestricted-injection sensitivity.}
The main Results report the small-commit maintainer simulation (Table~\ref{tab:kernel_results_summary}).
As a sensitivity check, we repeat the same one-per-author protocol without restricting the injected commit by size.
Table~\ref{tab:kernel_all_commit_sensitivity} shows that equal-mix raw then records the lowest median FP/pos (28), clearly ahead of the patch-size-gated dispatcher (69).

\OceanTabBegin
    \caption{Unrestricted one-per-author injection on the top-20 maintainer queue (sensitivity check).}
    \label{tab:kernel_all_commit_sensitivity}
    \begin{tabular}{@{}lrr@{}}
        \toprule
        \textbf{Rule} & \textbf{Med.\ FP/pos} & \textbf{Max FP} \\
        \midrule
        Text-only raw          & 82 & 1{,}385 \\
        Code-only raw          & 184 & 1{,}249 \\
        Equal mix raw          &  \textbf{28} & 251 \\
        Patch-size-gated dispatcher & 69 & 1{,}067 \\
        \bottomrule
    \end{tabular}
\OceanTabEnd

\subsection{PHP Case-Study Details}
\label{app:php_case_study_details}

Table~\ref{tab:php_results} expands the main PHP result with raw and percentile variants for every scoring rule. Percentile scores use prior-only mode, ranking each commit against the same author's earlier commits so that no future history leaks into the score. Both forged commits have a 637-character diff, below $T_{\mathrm{lo}}=1{,}000$.

\OceanTabBegin
    \caption{PHP maintainer-scoped detection (full rules).}
    \label{tab:php_results}
    \begin{tabular}{lrr}
        \toprule
        \textbf{Rule} & \textbf{FP before 1st} & \textbf{FP before all} \\
        \midrule
        Text-only raw (\texttt{text\_dist})          &     309 &     605 \\
        Code-only raw (\texttt{code\_dist})          & 11{,}700 & 13{,}622 \\
        Equal mix raw (\texttt{equal\_mix})          &  1{,}186 &  2{,}305 \\
        Patch-size-gated raw (\texttt{context\_gate\_raw})  &     273 &     653 \\
        \midrule
        Text-only percentile (\texttt{text\_pct})    &     549 &     598 \\
        Code-only percentile (\texttt{code\_pct})    & 12{,}358 & 18{,}335 \\
        \textbf{Patch-size-gated dispatcher (\texttt{context\_gate\_pct})} & \textbf{248} & \textbf{274} \\
        \bottomrule
    \end{tabular}
\OceanTabEnd

\subsection{ForceMemo Case-Study Details}
\label{app:fm_case_study_details}

The ForceMemo evaluation uses a curated subset of 40 repositories from the GlassWorm campaign (the full campaign affected hundreds of repositories~\citep{ForceMemoGlassWorm2026}).
Authors with fewer than five commits in the target repository are augmented with up to 30 genuine commits harvested from other repositories on GitHub (\emph{path-1 augmentation}), raising the scoreable spoof count from 22 to 28.
Of the 62 total spoof commits in the dataset, 34 remain unscoreable because impersonated authors had fewer than five prior commits available after three compounding effects: sparse histories in small target repositories, post-disclosure force-push cleanup by legitimate repository owners before our crawl, and one identity excluded by the Python-only file filter.
Commits are ranked within each repository. The top-$K$\% threshold is applied independently per repository.
Because all 28 scoreable GlassWorm payloads have $\ell \ge 6{,}664$ diff characters, the patch-size-gated dispatcher routes every commit to the code channel and selects raw cosine distance over per-author percentile rank (median prior count per compromised contributor is 15, below $T_{\mathrm{priors}}=100$).

Table~\ref{tab:fm_results} expands the main ForceMemo result with percentile variants and a maintainer-level breakdown restricted to the 11 repositories whose compromised contributor had at least 50 prior commits.

\textbf{Window choice.} Unlike the kernel and PHP detectors at $W=20$, the ForceMemo detector uses the smaller $W=5$. This is a coverage exception to the default $W=20$ transfer setting, not a validation-derived detection parameter. A commit is scoreable only if its author has at least $W$ prior commits in the cohort, so a smaller window scores more of each thin author's own history, which matters because the campaign targets sparse identities (the smallest scoreable cohort has roughly a dozen commits). The choice is governed by data coverage, not detection performance, and it is not selected by tuning against the spoof labels. Table~\ref{tab:fm_window_loss} bounds the cost of a larger window at the repository level: windows up to $W=10$ retain all $28$ affected repositories, $W=15$ begins dropping them, and $W=20$ discards $9$ of $28$, roughly a third of the corpus. The 11-repository maintainer block, whose contributors have ample history, is unaffected at every window. We keep the small operating point so that campaign breadth across sparse identities, the transfer condition this case is designed to test, is preserved.

\OceanTabBegin
    \caption{ForceMemo scoreability vs.\ detector window.}
    \label{tab:fm_window_loss}
    \begin{tabular}{ccc}
        \toprule
        $W$ & \textbf{All-cohort repos} & \textbf{Maintainer repos} \\
        \midrule
        \textbf{5} & \textbf{28} & \textbf{11} \\
        8  & 28 & 11 \\
        10 & 28 & 11 \\
        15 & 26 & 11 \\
        20 & 19 & 11 \\
        \bottomrule
    \end{tabular}
\OceanTabEnd

\textbf{Path-1 augmentation confounder analysis.}
Six of the 28 scoreable spoofs belong to authors whose in-target-repository history was below the five-commit minimum. Their baseline window was supplemented with commits from other GitHub repositories by the same contributor (\emph{path-1 augmentation}).
We split the 28 spoofs into augmented ($n=6$) and non-augmented ($n=22$) subgroups and compare Med.\ FP/pos for the code-only raw channel: non-augmented authors achieve Med.\ FP/pos $= 1$ (same as the full-corpus result), while augmented authors achieve Med.\ FP/pos $= 2$. Max FP is the same in both subgroups (Max FP $= 110$). Given the small augmented sample ($n=6$), this difference should be interpreted cautiously. Path-1 augmentation introduces a minor confounder that does not substantially change the headline result but may inflate individual-author FP counts for low-history contributors.

\OceanTabBegin
    \caption{ForceMemo detection results (28 scoreable spoofs). Parenthesized percentages express each spoof's false positives as a share of its own repository's scored queue, with the median and maximum then taken over these per-spoof shares. All seven rules reach their maximum in the same repository (BrianElionDev/BuyBot, queue size 300).}
    \label{tab:fm_results}
    \OceanResizeTable{
    \begin{tabular}{lrrrr}
        \toprule
        & \multicolumn{2}{c}{\textbf{All spoofs} ($N=28$)} & \multicolumn{2}{c}{\textbf{Maintainer-level} ($N=11$)} \\
        \cmidrule(lr){2-3}\cmidrule(lr){4-5}
        \textbf{Rule} & \textbf{Med.\ FP/pos} & \textbf{Max FP} & \textbf{Med.\ FP/pos} & \textbf{Max FP} \\
        \midrule
        Text-only raw (\texttt{text\_dist})   &   6 (21.4\%) &   262 (87.3\%) &    8 (8.6\%) &   262 (87.3\%) \\
        Code-only raw (\texttt{code\_dist})   &   1 (6.1\%) &   110 (36.7\%) &    1 (1.7\%) &   110 (36.7\%) \\
        Equal mix raw (\texttt{equal\_mix})   &   3 (7.7\%) &   243 (81.0\%) &    9 (6.7\%) &   243 (81.0\%) \\
        \textbf{Patch-size-gated dispatcher}         & \textbf{1 (0.8\%)} & \textbf{126 (42.0\%)} & \textbf{2 (0.5\%)} & \textbf{126 (42.0\%)} \\
        \midrule
        Text-only pct (\texttt{text\_pct})    &   9 (20.9\%) &   265 (88.3\%) &    9 (9.4\%) &   265 (88.3\%) \\
        Code-only pct (\texttt{code\_pct})    &   2 (5.6\%) &   111 (37.0\%) &    2 (1.7\%) &   111 (37.0\%) \\
        Mean pct (\texttt{mean\_pct})         &   3 (11.5\%) &   208 (69.3\%) &    2 (0.9\%) &   208 (69.3\%) \\
        \bottomrule
    \end{tabular}
    }
\OceanTabEnd

\textbf{Repository-membership caveat.}
Two repositories (\seqsplit{mystic3raven/python\_web\_app} and its sibling \seqsplit{mystic3raven/python-web-app-eksf}) have ranking queues larger than their individual commit counts. The repositories share overlapping history, so the repository-membership join attributes some shared commits to both queues. This caveat does not determine the reported Max FP, which occurs in \seqsplit{BrianElionDev/BuyBot}.

\textbf{Residual-case analysis.}
The heaviest review burdens are not evenly spread.
Under the dispatcher, 24 of the 28 repositories require reviewing at most $13.3\%$ of their scored queue (median under $1\%$), while the remaining four (\seqsplit{BrianElionDev/BuyBot}, \seqsplit{ThiagoGuerra09/TCC}, \seqsplit{AceLyo/Interval\_Timer\_Python}, and \seqsplit{DanielArmas22/clothing-shop-scraping}) require $30.8$--$42.0\%$.
These four cases are associated with a property of the impersonated account's own history rather than with payload size alone.
Their histories are dominated by bulk, heterogeneous drops or repetitive large edits, so a payload-bearing bulk commit is less exceptional against that baseline.
Quantitatively, the spoof's code-channel distance falls between the 38th and 63rd percentile of the impersonated account's own scored history in these four repositories, versus the 69th or higher (typically above the 85th) in the other 24 repositories below the $13.3\%$ burden cutoff.
The observed association is therefore with weak baseline separability (the spoof sits nearer the middle of the account's own distance distribution), not with file-type mix or commit size in isolation. This is a post-hoc reading of the residual cases, not a pre-registered predictor evaluated across the full campaign.